# Aqueous Proton Transfer Across Single Layer Graphene

Jennifer L. Achtyl[1], Raymond R. Unocic[2], Lijun Xu[4], Yu Cai[5], Muralikrishna Raju[6], Weiwei Zhang[6], Robert L. Sacci[3], Ivan V. Vlassiouk[3], Pasquale F. Fulvio[7,8], Panchapakesan Ganesh[2], David J. Wesolowski[8], Sheng Dai[8], Adri C. T. van Duin[6], Matthew Neurock[4,5], and Franz M. Geiger[1]*

[1]Department of Chemistry, Northwestern University, 2145 Sheridan Road, Evanston, IL 60201, USA.

[2] Center for Nanophase Materials Sciences Oak Ridge National Laboratory, Oak Ridge, TN 37831, USA.

[3]Measurement Science & System Engineering Division, Oak Ridge National Laboratory, Oak Ridge, TN 37931, USA.

[4]Departments of Chemical Engineering and Chemistry 102 Engineers' Way, University of Virginia, Charlottesville, VA 22904-4741, USA.

[5]Department of Chemical Engineering and Materials Science, 421 Washington Avenue S.E., University of Minnesota, Minneapolis, MN 55455.

[6]Department of Mechanical and Nuclear Engineering, Pennsylvania State University, University Park, PA 16801, USA.

[7]Department of Chemistry, University of Puerto Rico, Río Piedras Campus; San Juan, 00931 PR.

[8]Chemical Sciences Division, Oak Ridge National Laboratory, Oak Ridge, Tennessee 37831, USA.

*Correspondence and requests for materials should be addressed to F.M.G. (email: geigerf@chem.northwestern.edu).

**Abstract**

Proton transfer across single layer graphene is associated with large computed energy barriers and is therefore thought to be unfavorable at room temperature unless nanoscale holes or dopants are introduced, or a potential bias is applied. Here, we subject single layer graphene supported on fused silica to cycles of high and low pH and show that protons transfer reversibly from the aqueous phase through the graphene to the other side where they undergo acid-base chemistry with the silica hydroxyl groups. After ruling out diffusion through macroscopic pinholes, the protons are found to transfer through rare, naturally occurring atomic defects. Computer simulations reveal low energy barriers of 0.68 to 0.75 eV for aqueous proton transfer across hydroxyl-terminated atomic defects that participate in a Grotthuss-type relay, while pyrylium-like ether terminations shut down proton exchange. Unfavorable energy barriers to helium and hydrogen transfer indicate the process is selective for aqueous protons.

**Introduction**

Brick-and-mortar networks of stacked graphene oxide nanosheets can act as effective membranes[1-8] while single layer graphene exhibits dramatically lower permeabilities towards gases[4,9]. In fact, graphene is thought to be unfit even for proton transfer, which is associated with computed gas phase energy barriers exceeding 1.4 eV[10] unless dopants or nanoscale openings are externally introduced[6,7,10,11], or an external potential bias is applied[12]. To determine whether graphene is indeed impermeable to protons, we placed well-characterized single layer graphene[13] on top of a fused silica substrate and cycled, at room temperature and constant ionic strength, the bulk pH of an aqueous solution above the graphene layer between basic and acidic. We tested for proton exchange through graphene by probing the underlying silica surface with an interfacial potential-dependent version of second harmonic generation (SHG)[14,15] using 120 fsec input pulses at energies well below the graphene damage threshold[13]. With a detection limit of $10^{-5}$ to $10^{-6}$ V[16], the method is sensitive enough to follow protonation or deprotonation of as little as 1% of the available silanol groups present in the area probed by SHG. The interfacial potential vanishes at the point of zero charge (PZC of fused silica ~2.5)[17] and the SHG signal intensity is small[14,18,19]. Increasing the pH at constant ionic strength shifts the relevant interfacial acid-base equilibria $SiOH_2^+ + OH^- \rightleftharpoons SiOH + H_2O$ and $SiOH + OH^- \rightleftharpoons SiO^- + H_2O$ (pKa ~4.5 and ~8.5)[14,18,20] to the right and the resulting interfacial potential polarizes the interfacial water molecules such that the SHG signal intensity increases[14,18]. Intuitively, the close proximity of the graphene layer and the charged fused silica surface, combined with the sensitivity of the method, make our approach akin to an Å-scale voltmeter for detecting even rare occurrences of proton exchange. We find no significant difference between the SHG vs. time traces recorded in the presence and absence of graphene. After ruling out diffusion through macroscopic pinholes, the protons are found to transfer through rare, naturally occurring atomic defect sites. Computer simulations reveal low energy processes for water-mediated proton transfer across hydroxyl-terminated atomic defect sites that participate in a Grotthuss-type relay, while defects terminated by pyrylium-like ether bridges shut down proton exchange.

**Results**

**Silanol protonation and deprotonation unimpeded by graphene.**

Using a dual-pump flow system (Fig. 1a) at a flow rate of at ~0.9 mL s$^{-1}$, we varied the bulk solution pH between 3 to 10 while maintaining constant 1 mM ionic strength (see Methods). As shown in Fig. 1b, we find no significant difference between the SHG vs. time traces recorded in the presence and absence of graphene, and no statistically significant differences in the kinetic rates and jump durations (Supplementary Note 1). The SHG responses to pH changes are consistent with the acid-base equilibria of the fused silica/water interface[14,15,19,21], yielding effective $pK_{a,eff}$ values of 3.5(1) and 8.3(2), which fall within the reported literature values (Supplementary Note 2)[22]. This finding indicates that the SHG experiments do not track merely ion adsorption to the graphene/water interface but acid-base chemistry at the fused silica surface underneath it, for which proton transfer across the membrane is a necessary condition. As expected from refs 1-5, porous graphene multilayers do not inhibit proton transfer either (Supplementary Note 3). Based on these results we conclude that the fused silica/water interface does not behave differently in terms of relative surface charge density, in the duration of the jumps, or in the rates of the jumps when single layer graphene is present. These findings indicate that the acid/base chemistry at the fused silica/water interface occurs in an unimpeded fashion in the presence of single layer graphene.

**Importance of macroscopic defects ruled out.**

Scanning electron microscopy (SEM, Methods) images of graphene single layers deposited on fused silica windows show a low density of macroscopic pinholes and that the graphene is free of cracks or folds (Fig. 1c). Two-dimensional diffusion from those locations to the location of the laser beam is considered by calculating, for a given proton diffusion coefficient D, the mean-square displacement, $\langle \Delta r^2 \rangle$, according to $\langle \Delta r^2 \rangle = z \cdot t \cdot D$, where $t$ is time and $z$ is the number of neighboring sites to which the proton can hop[23] (six in for the case of the hexagonal graphene lattice). In the literature, reported theoretically and experimentally determined proton surface diffusion coefficients range between 1.01 x $10^{-7}$ cm$^2$ s$^{-1}$ and 9.00 x $10^{-5}$ cm$^2$ s$^{-1}$ [24-39]. While the bulk diffusion coefficient for a proton in water is accepted to range between 8 x $10^{-5}$ cm$^2$ s$^{-1}$ and 9 x $10^{-5}$ cm$^2$ s$^{-1}$, there are disagreements in the literature about whether the surface proton diffusion coefficient is similar to the bulk coefficient or slower than the bulk coefficient on hydrophobic and hydrophilic surfaces for a variety of different systems[24,28,38]. Reactivity is expected to substantially slow down the 2D diffusion of the proton (~ magnitude 20x reduction)[30,40,41] when it moves across an amphoteric oxide whose protonation effectively terminates the diffusion path. Reactive proton diffusion coefficients reported for Nafion[42,43] are similarly low. Indeed, our own reactive force field calculations containing partially hydroxylated quartz surfaces show the proton diffusion is quickly terminated by protonation of the surface $SiO^-$ groups (Supplementary Note 4). This result indicates that proton diffusion is significantly slower in the presence of surface anionic species due to proton trapping at these sites.

In our experiments, the continuous proton supply from the aqueous bulk is expected to form a propagating reaction front: our calculations show a drastically increased proton diffusion coefficient of $4.944 \times 10^{-5}$ cm$^2$ s$^{-1}$, or just half of that of bulk water, once protons arriving through any opening within the graphene sheet interact with the hydroxylated portion of the surface that is located behind the reaction front. To conservatively assess an upper bound limit for our estimations, we calculated the proton mean square displacement using a D value of $1 \times 10^{-6}$ cm$^2$ s$^{-1}$. The probability of placing our laser beam within the propagating reaction front emanating from a given macroscopic pinhole was then estimated to be 4 and 21% for 1 second and 10 second SHG jump times, respectively (Supplementary Note 5). Given that the pH jumps were repeated on least 18 different days with 8 different graphene samples and delays in changes of the SHG response were not observed with statistical significance, we conclude that the diffusion of protons from the few macroscopic pinholes that are present in our samples, or, alternatively, from the sample edge, to the area probed by the laser cannot explain our observations of proton transfer through graphene.

**Imaging rare atomic defects.**

Scanning transmission electron microscopy (STEM) was then used to search for atomic defects using annular dark field (ADF) STEM imaging at 60 kV (see Methods). The majority of the images show perfect six-fold symmetry in the position of the carbon atoms and vast areas that lack grain boundaries and atomic, or vacancy, defects (Fig. 1d). Nevertheless, similar to prior reports of atomic scale vacancy point defects[44,45], we find, albeit rarely, atomic defects (Fig. 1e). Unless hydrocarbons or heavy metal atoms[46] are present in graphene, defect formation due to electron beam-induced etching (as opposed to ion bombardment or oxidative etching)[47] of pristine CVD graphene at the energies employed here is unlikely. Rather, the rare defects we observe on occasion are more likely to originate from the synthesis process or cosmic rays, as the STEM experiments are carried out below the knock-on damage threshold for graphene[48], and the femtosecond laser pulses are attenuated below the onset of processes other than SHG[13]. Given a lower limit to the estimated defect-to-defect distance of ~0.1 μm[49] (while difficult to determine accurately from Raman spectroscopy, the actual distance is likely to be longer), we assess the probability of placing our laser beam within the propagating reaction front emanating from a given atomic defect to be 100%.

**Discussion.**

To elucidate the mechanisms for proton transfer, we discuss findings from density functional theory (DFT) calculations (Fig. 2) and ReaxFF reactive force field molecular dynamics (Fig. 3)[50,51] simulations (Methods). DFT simulations track the detailed changes in the electronic structure and quantify corresponding activation barriers as protons transfer from the water layer through the graphene interface and exit into solution on the opposite side of the surface. The ReaxFF simulations provide a larger scale representation of the interfaces and explicitly include dynamics.

We find that the main restriction for aqueous proton transfer through pristine, defect-free graphene is the energy required to push the proton through the center of an aromatic ring in the hydrophobic graphene layer as is shown in Supplementary Fig. 11. While the protons readily migrate in the solution phase above and below the graphene surface via proton shuttling, they are unable to pass through the hydrophobic graphene layer. The energy costs to desolvate the proton from the aqueous layer and drive it through the center of an intact aromatic ring within the graphene layer are quite high and result in an activation barrier that is over 3.8 eV.

The basal planes of pristine graphene can, and do, contain rare atomic-scale defect sites comprised of carbon atom vacancies, as was shown in Fig. 1e. Our calculations indicate that while the activation barrier for proton transfer through a single vacancy site is over 1.9 eV lower than that for transfer through the pristine graphene surface, it is still nearly 2.0 eV due to the small size of the vacancy and the hydrophobicity of the surface. The formation of di- and tri-vacancy sites increase the diameter of the opening in the graphene layer and reduce the barrier further to ~1.5 eV but this barrier is still too high to permit aqueous proton transfer at room temperature.

Removal of four carbon atoms in a central aromatic ring in the graphene layer leads to the formation of the quad vacancy (4V) site shown in Supplementary Fig. 13c. This site is comprised of six coordinatively unsaturated carbon atoms that are either terminated with three oxygen atoms in epoxide-like arrangements reminiscent of pyrylium cations (different from the crown ethers recently reported by Guo et al.)[53], or with six hydroxide groups. All of the defect terminations considered are energetically favorable as compared to the bare quad-vacancy system (Supplementary Note 6). Proton transfer through the pyrylium-terminated 4V site requires 1.7 eV (Fig. 2a), attributed to the protophobicity of pyrylium cations and their in-plane localization, which leaves a 3.4 Å gap between water and the graphene substrate that prevents proton transfer. The hydroxyl-terminated site (Fig. 2b), however, provides hydrogen-bonding networks (Fig. 2d) that interconnect the graphene surface to the water layers above and below it. DFT and ReaxFF simulations indicate that these hydrogen-bonding networks serve as conduits that facilitate proton transfer from the solution phase to the surface through the center of the defect site and into the solution on the opposite side of the membrane via a Grotthuss mechanism[52] involving proton shuttling. This proton transfer mechanism identified here involves relaying the proton from one of the top three defect hydroxyl groups to the next hydroxyl group and the next, subsequent transfer to one of the bottom three defect hydroxyl groups on the other side, and finally release into the aqueous phase. While solution phase proton shuttling occurs with activation barriers < 0.2 eV, the barrier for transferring the proton through the defect sites in graphene via the proton relay mechanism is just 0.68 eV (DFT, Fig. 2b, well-reproduced by ReaxFF (0.61 eV)), indicating proton transfer will occur at room temperature.

Additional ReaxFF simulations show that a water channel, which establishes itself upon proton transfer, thins and finally vanishes when the pairs of OH groups terminating the defect site are successively replaced with oxygen atoms (Fig. 3a-d). These transfer paths are selective to aqueous protons as helium and $H_2$ transfer requires barriers exceeding 1.9 eV (Supplementary Note 7). Table 1 gives the comparison of activation barriers for proton transfer through graphene in water calculated by ReaxFF and DFT. The barriers given by DFT for the pristine and 1V case are high (3.9 eV and >2.0 eV respectively) and insurmountable during MD simulations at 300 K. ReaxFF overpredicts the barriers for proton transfer in the pristine and single vacancy (1V) case.

Yet, the barriers for the relevant 4V cases given by ReaxFF are in good agreement with DFT. Note that ReaxFF was not specifically trained against any of these barriers.

We conclude that aqueous protons transfer through single layer graphene via rare, OH-terminated atomic defects at room temperature. While the rarity of the atomic defect sites would make it challenging to follow proton exchange across graphene using pH-sensitive electrodes, the close proximity of the graphene layer and the charged fused silica surface, where the experimental observation of surface protonation and deprotonation is made by SHG, allows for the experimental observation of proton exchange across these rare defects. The associated energy barriers are comparable to recent experimentally determined activation energy barriers for proton transfer through graphene subjected to an externally applied potential[12]. From the SHG signal jump rates and the times required for 2D proton diffusion, we estimate that the presence of as few as a handful of atomic defects in a 1 µm$^2$ area sample of single layer graphene is sufficient to allow for the apparent unimpeded protonation and deprotonation of the interfacial silanol groups within ten seconds (Supplementary Note 8). Yet, we caution that given the limited accuracy with which the defect density can be determined in large (mm)-scale graphene, aqueous protons may transfer across single layer graphene not only along the path discussed here, but others as well. The identification of low barriers specifically for water-assisted transfer of protons through OH-terminated atomic defects in graphene, and high barriers for oxygen-terminated defects could be an important step toward the preparation of zero-crossover proton-selective membranes.

**Methods.**

**CVD graphene synthesis.** We used graphene having a grain size of ~100 µm[54] grown on copper foils by atmospheric pressure CVD[54,55]. The graphene was transferred using spincoating of poly(methyl methacrylate) (PMMA) followed by copper etching in a FeCl$_3$ solution and PMMA removal in acetone. The transfer was made onto clean fused silica substrates (ISP Optics, 1" diameter, QI-W-25-1, flatness 1 wave per inch at 633 nm) to fill approximately one cm$^2$ with a single layer. Following annealing in a flow of 4% H$_2$ in Ar for 30 min at 300 ºC, vibrational sum frequency generation spectra showed no evidence for CH stretches[56]. Similar to the finding of water layers between graphene and mica by atomic force microscopy[57], there is probably water located between the graphene samples and the fused silica substrates used here.

**Scanning Electron Microscopy (SEM)**. SEM images were collected from the center of the graphene film. Graphene on an optical window was imaged using a Hitachi S-4800 scanning electron microscope (SEM) operating at 2 kV with a probing current of 10 µA and an Everhart-Thornley detector. Copper tape was used to reduce charging effects. Individual images were taken at 1200x magnification with 1280 x 960 resolution. An array of 5x5 images (529 x 397 µm, pixel size 176 nm) was stitched together using Adobe Illustrator. Automatic brightness and contrast adjustment on each frame was carried out using the "auto adjust" feature in Preview (Apple, Inc.). No other post-edit feature or change was applied.

**Aberration-Corrected Scanning Transmission Electron Microscopy (STEM)**. To confirm the single layer nature of graphene synthesized using the CVD method[58], atomic resolution STEM imaging was performed at room temperature with an aberration-corrected Nion UltraSTEM-100[59] equipped with a cold field-emission electron source. The microscope was operated at 60kV, which is below the knock-on damage threshold for graphene. The CVD prepared graphene specimens were transferred to a SiN supported silicon microchip TEM grid. Prior to STEM imaging, the specimen was heated at 160º C in vacuum (10$^{-5}$ torr) for 8 hours to remove surface contamination. Following heating in vacuum, the specimen was immediately transferred to the UltraSTEM for ADF STEM imaging. The surface of the graphene still retains residual PMMA that was used in the transfer processes to the TEM grid as shown in Supplementary Fig. 26; however, there are large areas that are devoid of the PMMA which made it feasible to directly image the lattice structure and confirm the single layer nature using atomic resolution STEM imaging. The images were filtered using a smoothing function in Digital Micrograph and the contrast and brightness were adjusted to enhance the contrast of the graphene.

**Aqueous solution and substrate preparation.** The aqueous solutions were prepared with Millipore water, prepared the day prior to an experiment and left open to air overnight to equilibrate with atmospheric $CO_2$, and NaCl (Alfa Aesar, 99+%). The concentration of NaCl was confirmed using a conductivity meter (Fisher Traceable Conductivity and TDS meter, Fisher Scientific). Solution pH was adjusted with minimum amounts of dilute solutions of ~1M NaOH (Sigma-Aldrich, 99.99%) and HCl (EMD ACS grade). The pH jump experiments were carried out using a fused silica hemisphere (ISP Optics, 1" diameter, QU-HS-25) pressed against either a fused silica window (ISP Optics, 1" diameter, QI-W-25-1), or a CVD prepared graphene film transferred onto a silica window in an experimental setup previously reported[56,58]. The hemisphere and fused silica window were cleaned prior to experiments by first treating the surface of interest with NoChromix (Godax Laboratories) for 1 hour, rinsing with Millipore water and then storing in Millipore water overnight for SHG experiments the next day. The day of the experiment, the bare silica window and hemisphere were sonicated in methanol for 6 minutes, dried in a 110°C oven for 30 minutes, oxygen plasma cleaned (Harric Plasma) on high for 30 seconds, and then stored in Millipore water until the experiment. The graphene samples were not cleaned with this procedure, but were instead cleaned by flushing with approximately 2 L of Millipore water before each experiment. Supplementary Note 9 describes the graphene characterization and analysis by Raman and UV-vis spectroscopy prior and after the pH jump experiments.

**Flow system and flow cell.** As shown in Fig. 1a, the graphene-on-fused silica sample or the silica window were clamped face down against a Viton O-ring on the Teflon flow cell[13,56] so that the surface of interest was in contact with the aqueous phase. The fused silica hemisphere was then clamped on top of the window with a Millipore water layer in between in order to minimize the change of refractive index between the phases and to avoid use of an index-matching fluid. Throughout the duration of the experiment it was also necessary to maintain a ring of Millipore water around the bottom of the hemisphere in order to avoid evaporation of the sandwiched water layer. All of the experiments were completed with a 0.9 mL s$^{-1}$ flow using variable flow peristaltic pumps as previously reported[56,58]. Using the flow system depicted in Fig. 1a, the pumps were switched to pull solutions from two different reservoirs. For the experiments reported here (excluding pKa experiments, see Supplementary Note 2), the two reservoirs contained 1 mM NaCl Millipore solutions adjusted to either pH 3 or pH 10. At the start and end of each pH jump experiment, a 1 mM NaCl aqueous solution adjusted to pH 7 was pumped through the system and the SHG signal was collected until it reached steady state. It is assumed that steady-state conditions were reached once the SHG signal remained at a stable intensity for a minimum of 300 s. After the system reached steady-state at pH 7, the flow was switched back and forth between the pH 3 and pH 10 aqueous solutions, each time allowing the SHG signal to reach steady-state before switching to the next pH. After several pH 3 to 10 and pH 10 to 3 jumps were completed, the pH was adjusted back to pH 7, and the SHG signal was collected until steady-state was reached one last time. None of the liquid flow effects reported for fused silica/water interfaces subjected to high shear rates[60] were observed under the creeping flow conditions used here. Supplementary Note 10 assesses the flow dynamics in the cell.

**Laser and detection system.** A detailed description of our SHG setup has been described previously[61-64]. Briefly, we use a regeneratively amplified Ti:sapphire system (Hurricane, Spectra Physics) that operates at a kHz repetition rate to produce 120 femtosecond pulses to pump an optical parametric amplifier (OPA-CF, Spectra Physics) tuned to produce 600 nm light. After exiting the OPA, the beam is then directed through a variable density filter to attenuate the pulse energy to either $0.3 \pm 0.05$ µJ pulse$^{-1}$ for bare silica studies, or $0.15 \pm 0.05$ µJ pulse$^{-1}$ per pulse for graphene studies. The pulse energy used for the graphene films equates to a power density of $2.1(7) \times 10^4$ µJ cm$^{-2}$ per pulse with a 30 µm focal spot, which is well below the damage threshold of graphene as previously reported[56,58]. At an angle just below total internal reflection, the p-polarized attenuated fundamental light is then directed through a fused silica hemisphere and focused at the graphene/water or silica/water interface. The reflected fundamental and second harmonic lights are directed through a Schott filter and a monochromator to remove the any contributions at the fundamental frequency before amplification with a PMT and detection using a gated single photon counting system. Correct power dependencies and spectral responses are verified regularly, the SHG responses are well polarized, and sample damage does not occur[56,58]. Given that the SHG jump rates are independent of the mean stream velocity (Supplementary Note 1), we are confident that the acid/base reactions occurring at the fused silica surface are not mass transfer limited. UV-vis and Raman spectra shown in the Supplementary Information indicate that the samples are resistant to acid-base cycling under the conditions employed here.

**Computer simulations.** First principles periodic density functional theory calculations were carried out to determine the lowest energy interfacial water/graphene, water/graphene/water/silica structures and the activation barriers for proton diffusion through these interfaces using the Vienna Ab Initio Simulation Package (VASP)[65,66]. In the DFT calculations, the reaction systems were modeled by optimizing a water phase above and below a single graphene sheet. The simulations were carried out in a 5×5 supercell comprised of 50 carbon atoms, extended infinitively in the x and y dimensions. A 15 Å gap was inserted between the graphene layer perpendicular to the surface. The gap was subsequently filled with enough water molecules to match the overall density of water at $1.0\times10^3$ kg m$^{-3}$. The initial simulations were carried out with water on both side of the graphene layer. The lower $SiO_2$ substrate was initially simplified by using additional water. Subsequent calculations were carried out with more realistic slabs comprised of water/graphene/water/$SiO_2$ substrates. The reaction rates and mechanisms of proton transfer through the graphene were described in the framework of transition state theory and within the harmonic approximation, which is robust for systems of high densities.

All of the calculations were carried out within the Generalized Gradient Approximation using Perdew-Burke-Ernzerhof (PBE) functional[67] to treat exchange and correlation gradient corrections and PAW pseudopotentials[68] to describe the electron-ion interactions. Plane wave basis sets with a cutoff energy of 400 eV were used to solve the Kohn–Sham equations for calculations for systems without water. Calculations for systems that include water solvation were carried out with cutoff energies for C and O of 283 eV. The surface Brillouin zone was sampled using a Monkhorst-Pack mesh of 3×3×1. All electronic energies were converged to within a tolerance of $1\times10^{-5}$ eV. All of the atoms were allowed to relax in the geometry optimizations until the forces on each atom were less than 0.03 eV Å$^{-1}$. Spin polarization was examined for all of the systems explored and applied when needed. Transition states were isolated using the nudged elastic band (NEB) method[44,45] together with the dimer method[69]. The NEB method was used to provide an initial transition state structure that was used in the subsequent dimer simulations to isolate the transition state. The reaction barrier was defined as the energy difference between the transition state and the reaction state minimum. The intrinsic barrier is defined as the energy gap between a transition state and its immediate reaction state. Given the importance of surface relaxation in atomically defected graphene layers[70], all of our calculations on the 1, 2, and 4V carbon vacancy sites and the oxygen-terminated sites explicitly modeled surface relaxation (Supplementary Note 11).

The ReaxFF simulations were performed using the stand-alone ReaxFF implementation to study proton transfer through pristine graphene and graphene with di- and quad- vacancies. We then compared to results from long *ab initio* molecular dynamics (AIMD) to validate predictions of force field in describing water/graphene systems (Supplementary Note 12). In our simulations we used a (6x6) periodic graphene sheet with water molecules placed in random configurations on either side of the graphene sheet. The dimensions of the simulation cell are 15.01 Å x 17.83 Å parallel to the sheet and 30 Å in the direction perpendicular to the sheet. All MD simulations have been performed in the canonical (NVT) ensemble, with a time step of 0.25 fs using the Berendsen thermostat with a coupling time constant of 100 fs to control temperature of the entire system. To obtain the density plots in Fig. 3, we first divided the simulation cell into a mesh of cubic boxes with dimensions (0.30 Å x 0.30 Å x 0.30 Å). We then counted the number of times a particular atom type (*e.g.*, oxygen) was located in each of the grids through the entire length of simulation and normalized these numbers by the highest count recorded in any of the grids. We used these normalized values to obtain the resulting density plots in Fig. 3.

**Acknowledgments:** This work was supported by the Fluid Interface Reactions, Structures and Transport (FIRST) Center, an Energy Frontier Research Center funded by the U.S. Department of Energy, Office of Science, Office of Basic Energy Science. Microscopy conducted as part of a user proposal at ORNL's Center for Nanophase Materials Sciences (CNMS), which is an Office of Science User Facility. We also gratefully acknowledge helpful discussions with Wesley R. Burghardt.

**Author Contributions.** F.M.G. conceived of the idea. J. L. A., R. R. U., R. L. S., I. V. V., and P. F. F. performed the experiments. L. X., Y. C., M. R., W. Z., P. G., A. C. T. v. D. and M. N. performed the computational work. The manuscript was written with substantial contributions from all authors.

**Additional Information**

**Competing financial interests.** The authors declare no competing financial interests.

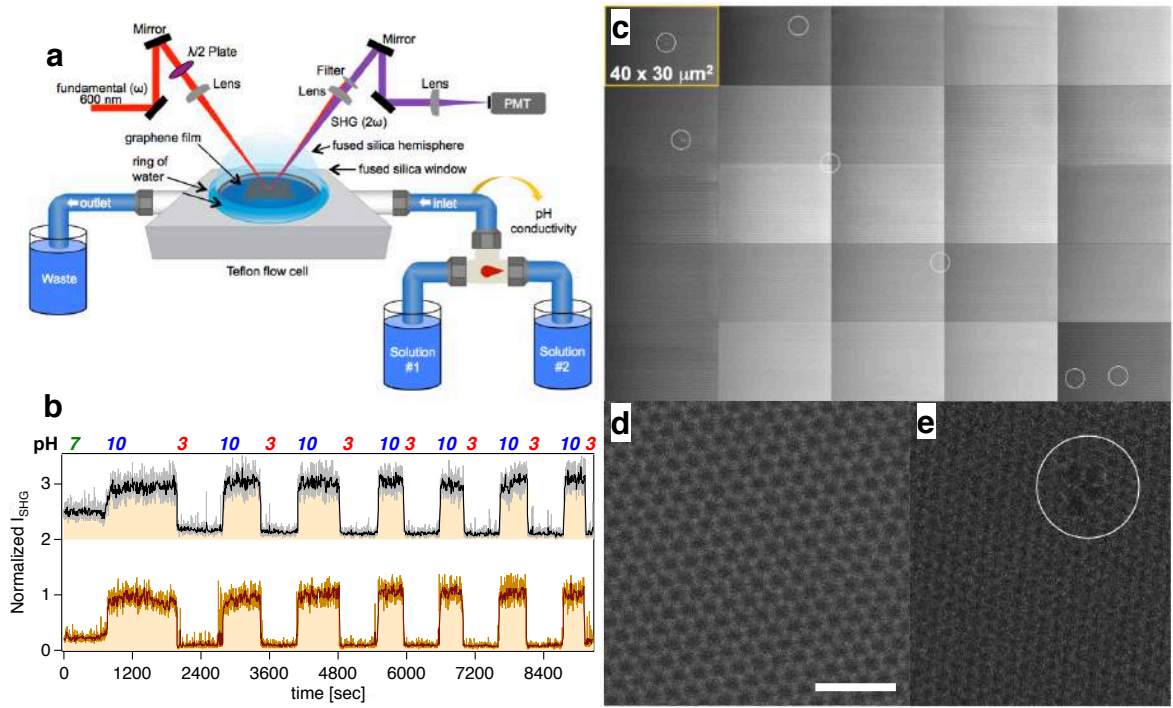

**Figure 1 | Experimental approach.** (a) Experimental setup using a waveplate (λ/2) to prepare 600 nm light plane-polarized parallel to the plane of incidence (p-in) while a photomultiplier tube (PMT) detects the second harmonic generation (SHG) photons at λ=300 nm. (b) p-in/all-out Polarized SHG response recorded as a function of time from the fused silica/water interface during pH jumps from 7 to 3 to 10 and subsequent pH cycling between 3 and 10 at a bulk aqueous flow of 0.9 mL/sec and 1 mM NaCl concentration in the absence (crimson, bottom) and presence (black, top, offset for clarity) of single layer graphene placed between the fused silica substrate and the flowing bulk aqueous phase. 5-Point boxcar indicated by dark lines. (c) Composite of 25 SEM images of single layer graphene on a fused silica substrate, showing 7 macroscopic pinholes, marked by white circles. (d) High resolution aberration-corrected ADF STEM

images of defect-free single layer graphene on a TEM grid and **(e)** of a rarely imaged atomic defect.

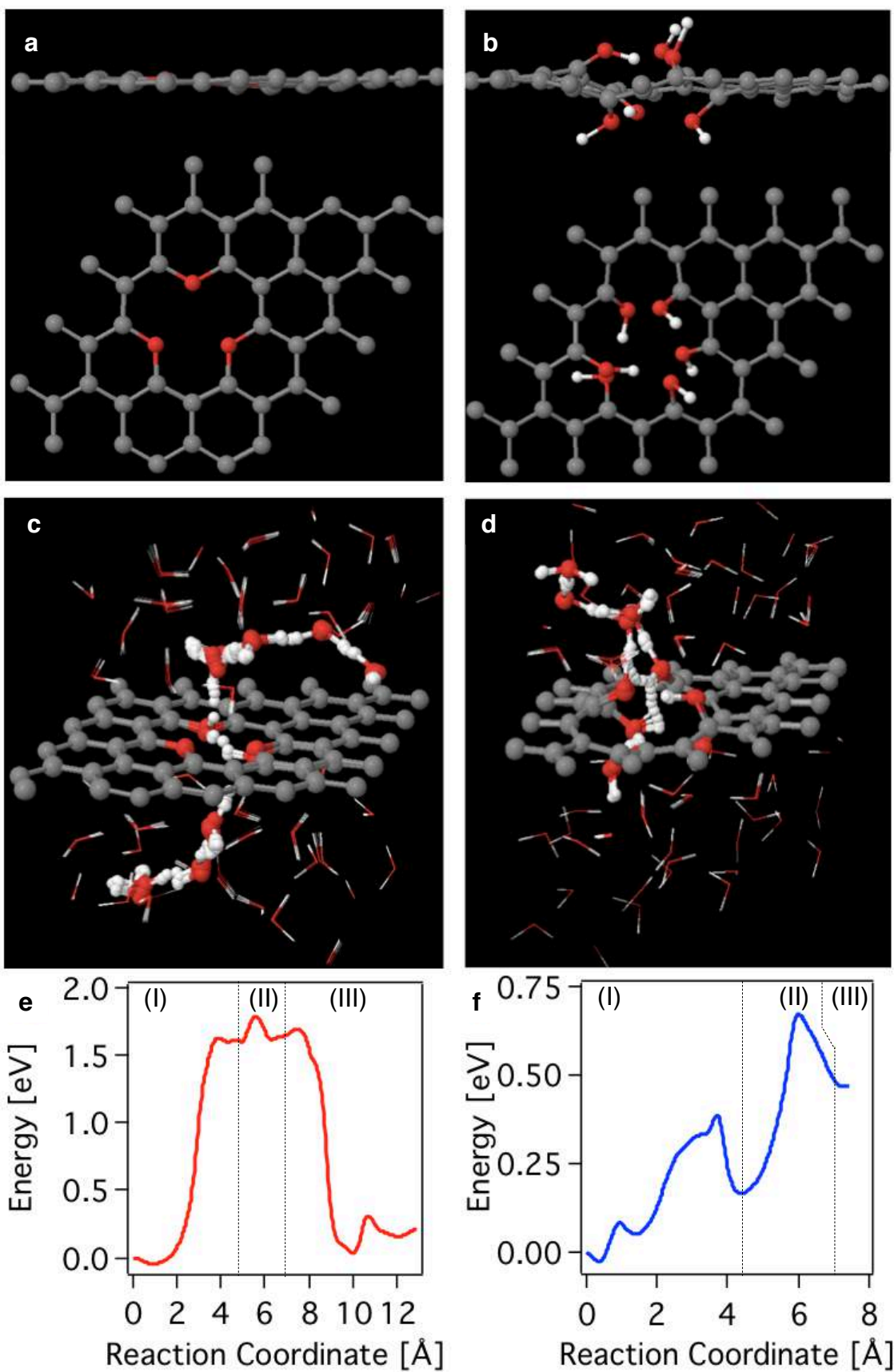

**Figure 2 | Density Functional Theory calculations.** Side and top views of oxygen- (**a**) and OH-(**b**) terminated defect models used in the DFT calculations. Snapshots (**c**, **d**) and energetics (**e**, **f**) from the nudged elastic band calculations for proton transfer through the oxygen- and OH- terminated defect sites marking (**region I**) release of proton from $H_3O^+$ to oxygen and OH group, respectively; (**region II**) relay of proton between oxygen and OH groups, respectively; (**region III**) release of proton from oxygen and OH groups to $H_3O^+$, respectively. Denotations of spheres: grey=carbon; red=oxygen; white=hydrogen atoms.

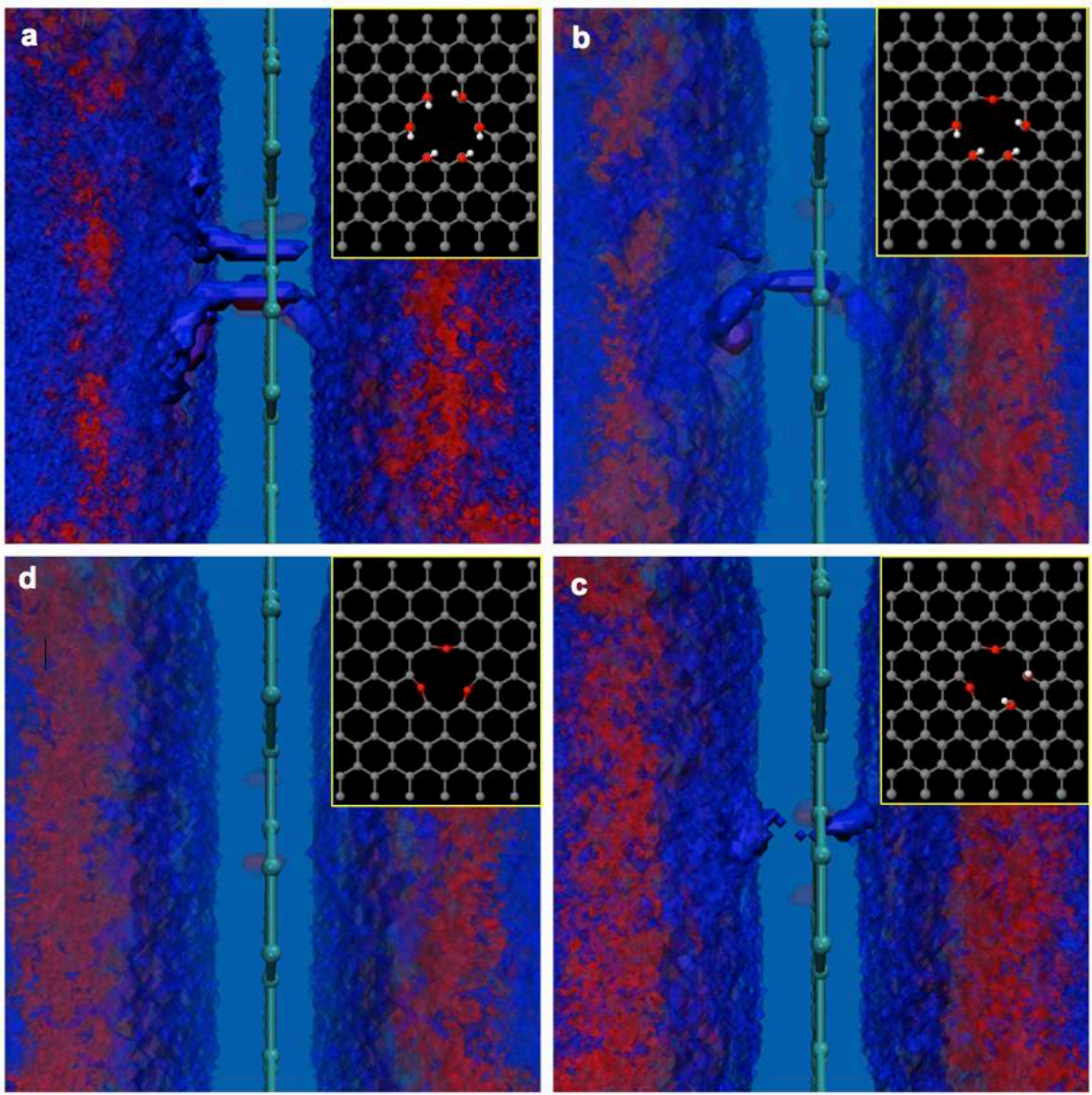

**Figure 3 | Reactive Force Field calculations.** Water channel formation from ReaxFF calculations of water mediated proton transfer through atomic defects terminated in 6 OH groups **(a)**, 4 OH groups and one oxygen atom **(b)**, 2 OH groups and two oxygen atom **(c)**, and three oxygen atoms **(d)**. Denotations of spheres: grey=carbon; red=oxygen; white=hydrogen atoms.

**Table 1**

**DFT and ReaxFF-calculated activation barriers for proton transfer through different vacancy sites on graphene in water**

| Graphene Surface | Bottom Layer | Defect Termination | Activation Barrier DFT* | Activation Barrier ReaxFF* |
|---|---|---|---|---|
| No vacancy | water | No termination | 3.9 eV | Not computed |
| 1V | water | No termination | > 2.0 eV | 3.54 eV |
| 4V | water | No termination | 0.25 eV | 0.22 eV |
| 4V | water | 3O ether capped | 1.8 eV | 1.7 eV |
| 4V | water | 6OH hydroxyl capped | 0.68 eV | 0.61 eV |
| 4V | Water + $SiO_2$ | 3O ether capped | 2.5 eV | 2.53 eV |
| 4V | Water + $SiO_2$ | 6OH hydroxyl capped | 0.7 eV | 0.75 eV |

*The energy difference that is reported is due to the initial reference (or reactant) state.

**Supplementary figures**

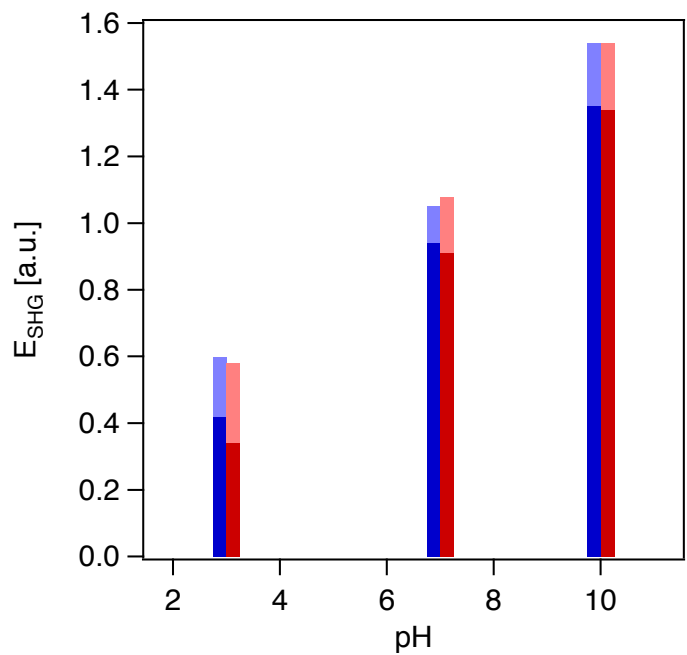

**Supplementary Fig. 1 | pH-dependent SHG E-fields from fused silica with and without graphene.** Comparison of the average SHG E-fields from pH 3, pH 7, and pH 10 adjusted 1 mM NaCl Millipore solutions over a single layer graphene film (blue bars) and a bare fused silica window (red bars). The resulting SHG E-fields were normalized to the averaged E-field calculated from the SHG intensities collected from a pH 7 solution at the beginning and end of each "pH jump experiment". Lighter red and blue colors represent the ± uncertainties (1σ) of the point estimates.

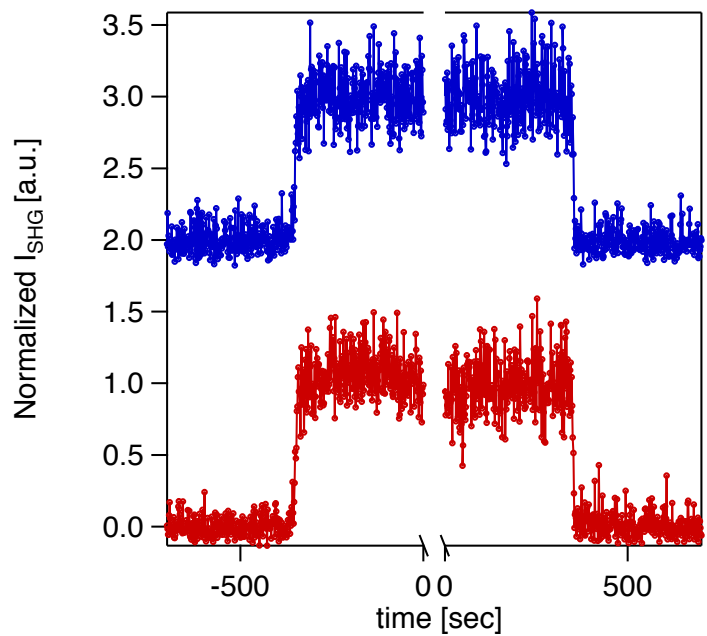

**Supplementary Fig. 2 | SHG vs time traces.** A comparison of the duration of the 3 to 10 pH jumps (left) and 10 to 3 pH jumps (right) for the single layer graphene film (blue traces) and the bare fused silica (red traces) at a 0.3 mL/sec flow rate. Each pH jump time trace is referenced to pH 3 and normalized to pH 10 and is the average of 4 individual SHG pH jumps. The different SHG traces were averaged together centering each jump such that the calculated inflection points were centered at the same time value. The graphene traces are offset for clarity.

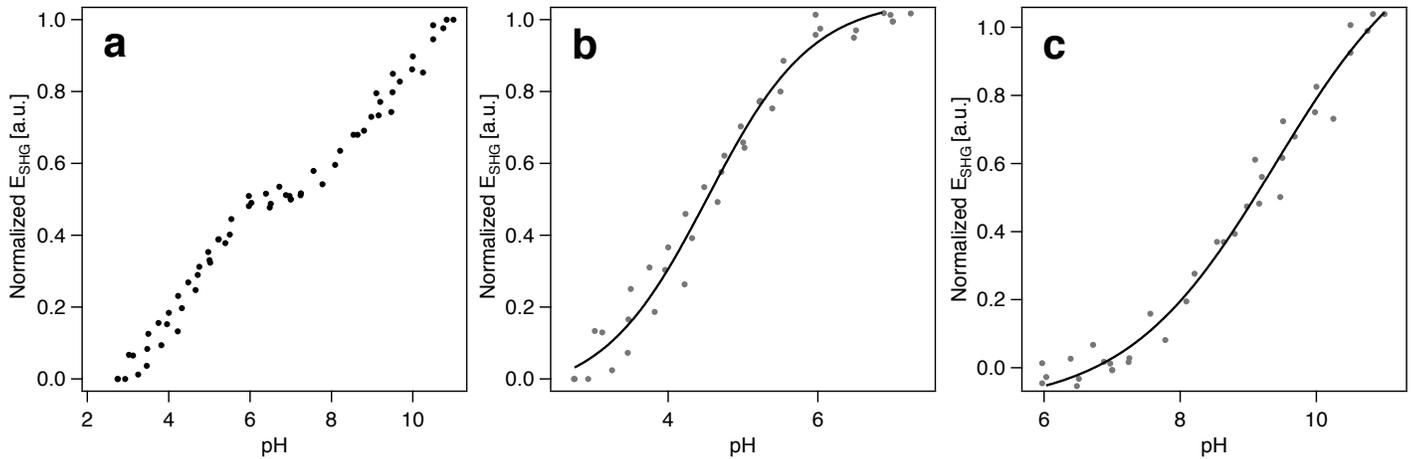

**Supplementary Fig. 3 | Interfacial titration curves by SHG.** The normalized SHG E-field plotted as a function of pH from the water/single layer graphene/fused silica interface in the presence of 100 mM NaCl. The data is the compilation of 3 separate experiments run on 2 different graphene samples. The E-field values from each day are normalized to the averaged E-fields obtained from pH 7 aqueous solutions. **(a)** The SHG E-fields referenced to the minimum E-field and normalized to the maximum E-field, showing the two inflection points consistent with the bimodal acid-base equilibria of the fused silica interface. **(b)** The SHG E-fields referenced to the E-field value at pH 2.75 and normalized to E-field value at pH 7.25. The line represents the sigmoid fit of the data. The inflection point here was used to calculate the $pK_a^{eff}$ of the more acidic silanol groups. **(c)** The SHG E-fields referenced to the E-field value at pH 6 and normalized to E-field value at pH 11. The line represents the sigmoid fit of the data. The inflection point here was used to calculate the $pK_a^{eff}$ of the less acidic silanol groups.

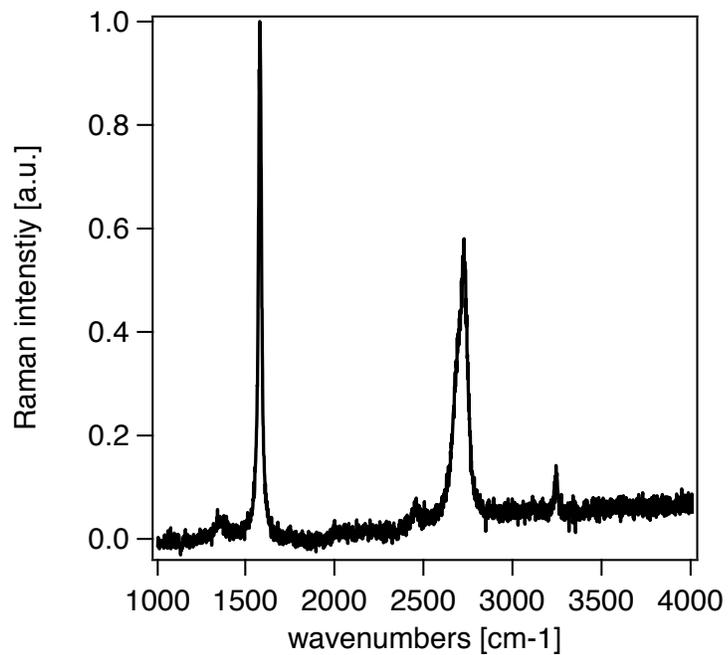

**Supplementary Fig. 4 | Raman spectra of multilayer graphene.** A representative Raman spectrum obtained from the multilayer graphene sample with the characteristic 2D, G, and D bands indicative of multilayer graphene films. The spectrum is normalized to the maximum Raman intensity value.

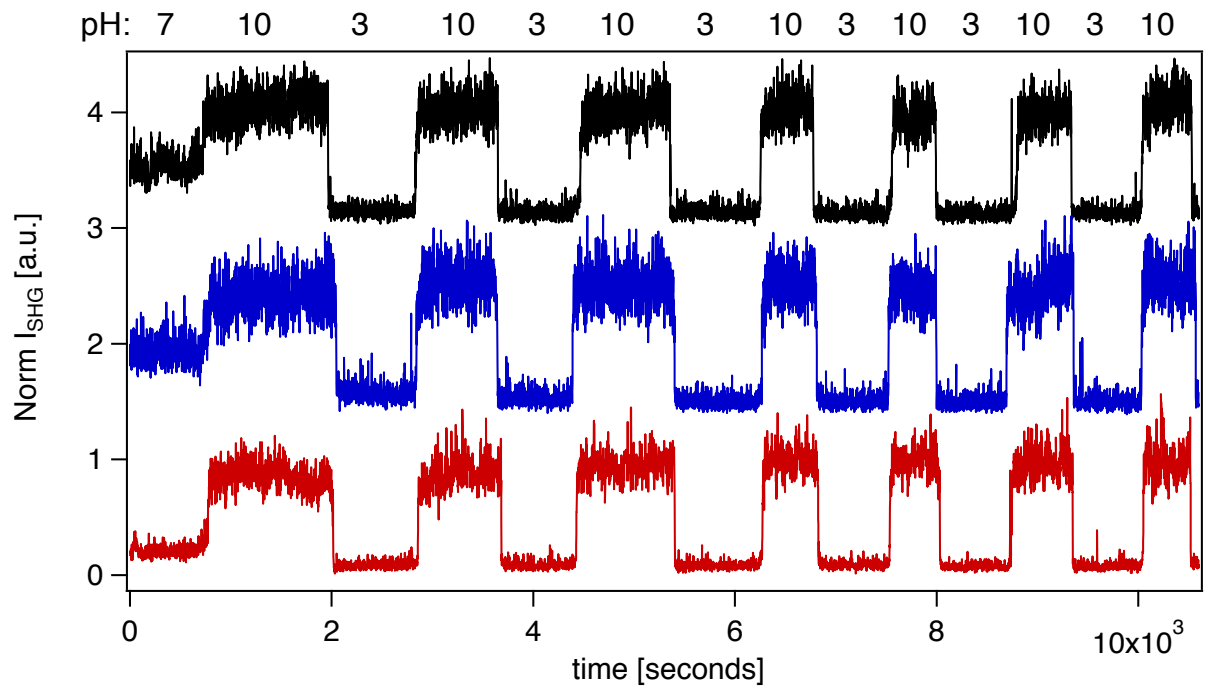

**Supplementary Fig. 5 | SHG vs time traces.** Normalized SHG intensity recorded as a function of time from the water/bare fused silica (crimson trace), water/single layer graphene/fused silica (blue trace) and water/multilayer graphene/fused silica (black trace) interfaces during "pH jumps experiments". The traces were recorded with a 0.9 mL/sec flow rate starting at pH 7 and then cycling between pH 10 and 3 with a 1mM NaCl Millipore water solution.

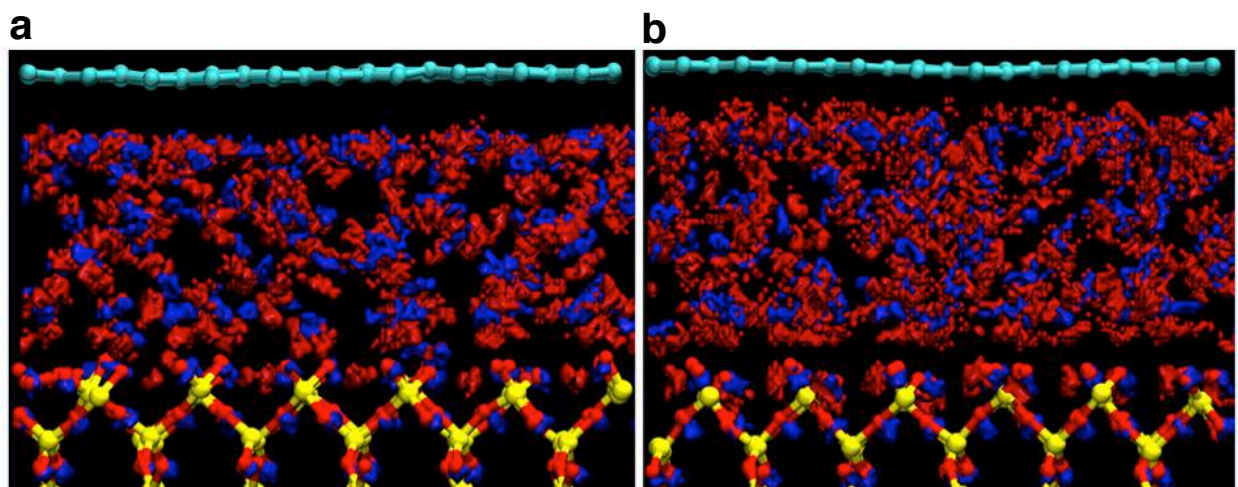

**Supplementary Fig. 6 | Water structure between $SiO_2$ and graphene.** Density plot of oxygen(blue) and hydrogen(red) in the silica/water/graphene interface, as obtained during a 300K ReaxFF molecular dynamics simulation. (**a**) structure at high pH (40% SiOH/60% SiO$^-$; OH-ions in water solution) – showing very strong local water structure and 20-fold reduced water self diffusion (**b**) water structure at neutral/low pH (100% SiOH; pure $H_2O$) showing more liquid water structure and water self-diffusion comparable to bulk water. Denotations of spheres: yellow=silicon; red=oxygen; cyan=carbon atoms.

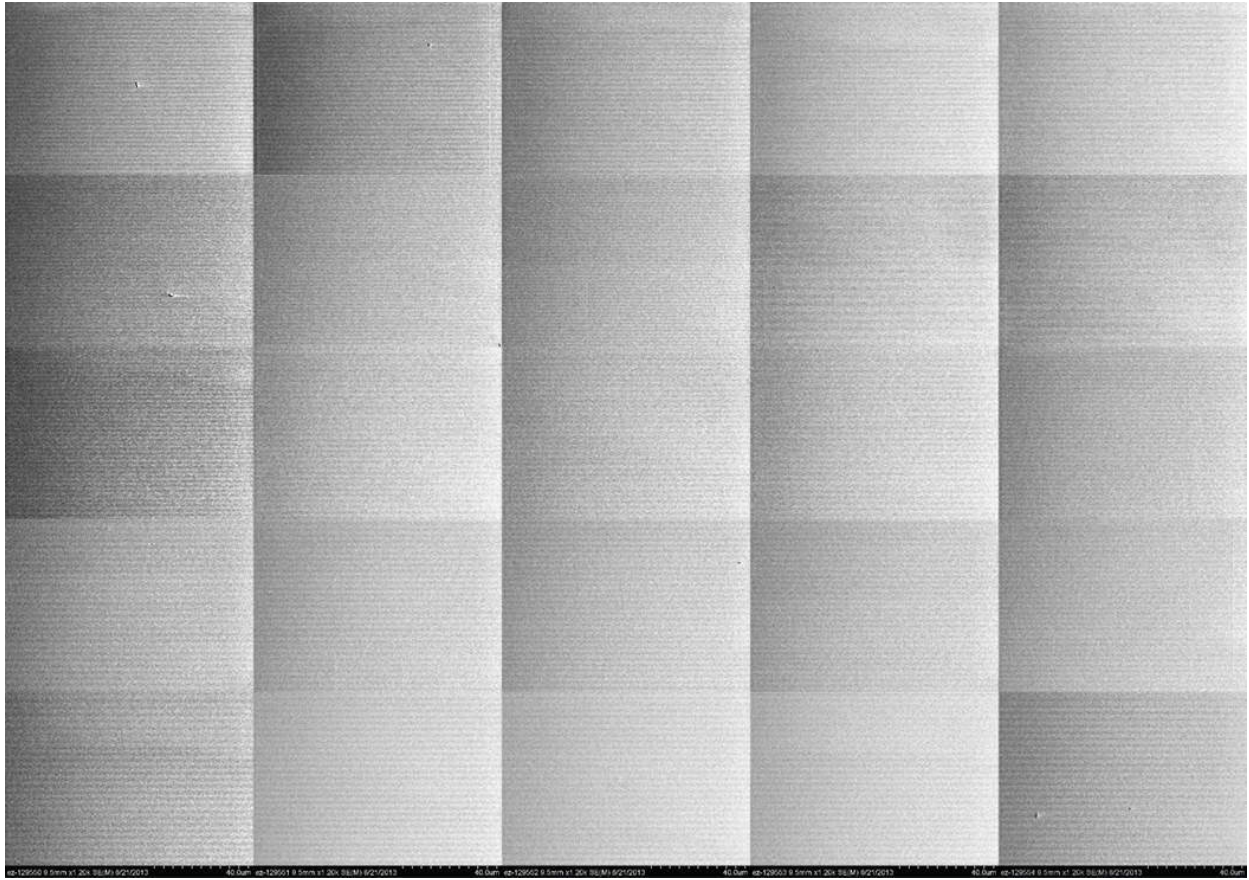

**Supplementary Fig. 7 | Imaging macroscopic defects by SEM.** Resulting image from combining the SEM images collected over a 529 $\mu$m x 397 $\mu$m area at the center of the graphene film.

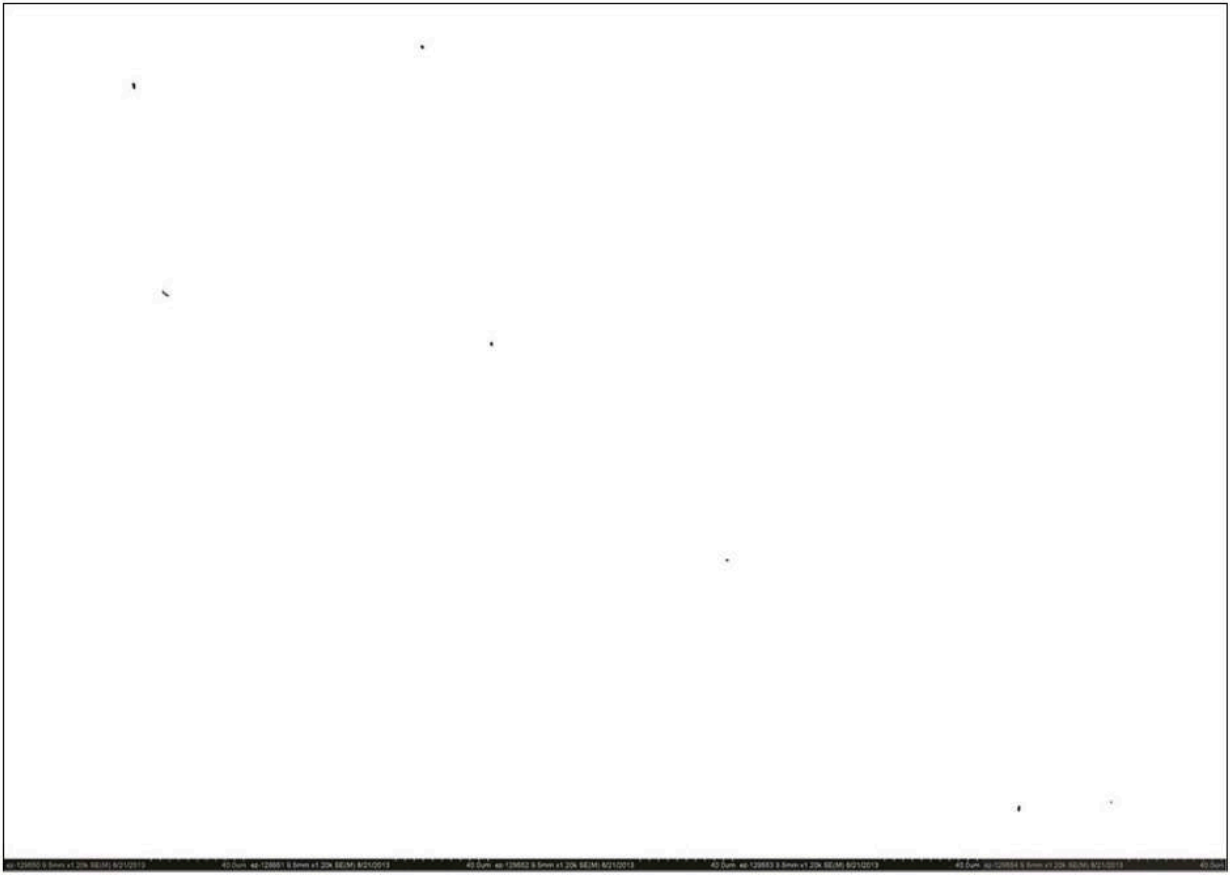

**Supplementary Fig. 8 | Locating macroscopic defects.** The resulting "masks" image returned from running the ImageJ "Analyze Particles" program after thresholding on the pinhole-marked SEM image as described in the Supporting Information text. The dark features represent the detected anomalies larger than 9 pixels in size. The 7 anomalies shown here were all counted and treated as pinholes.

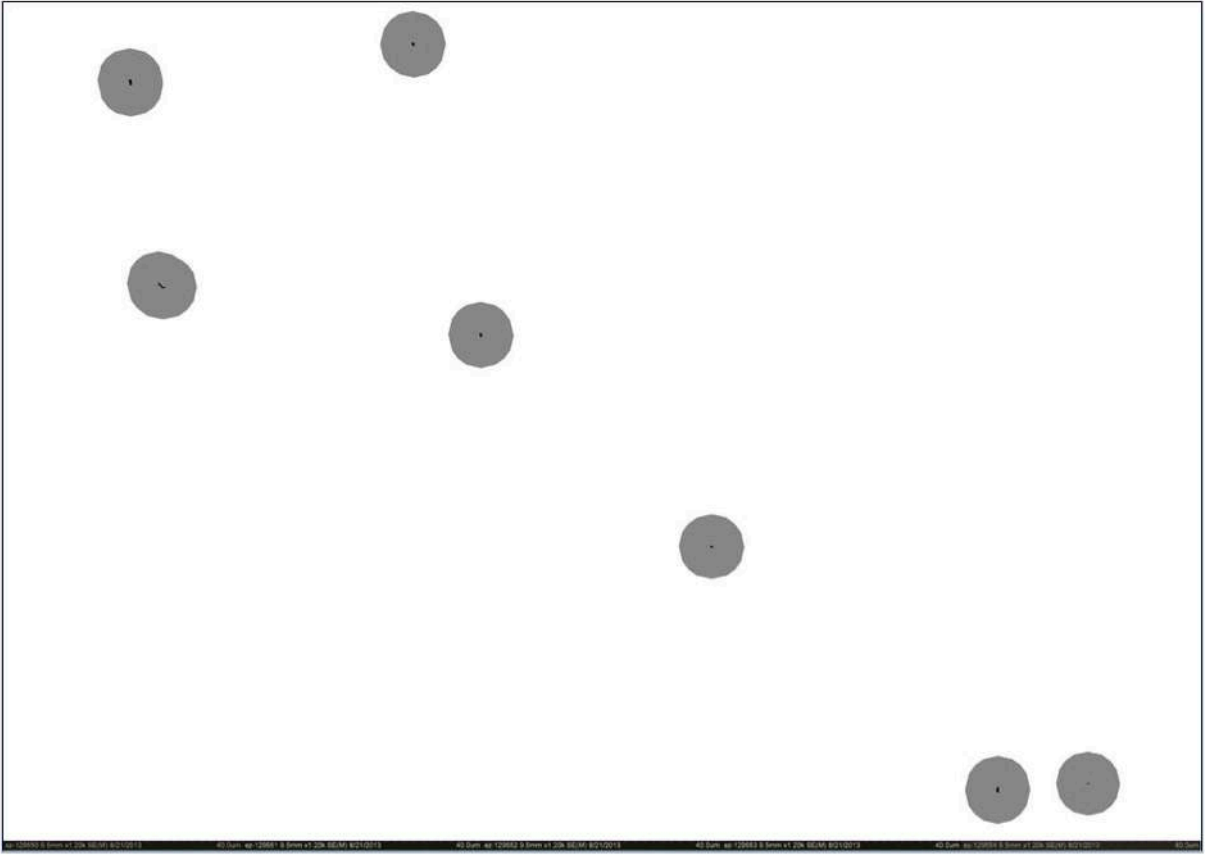

**Supplementary Fig. 9 | 2D-diffusion calculation for macroscopic defects I.** The resulting image after increasing the size of the pinholes (black) on all sides by the proton's radius of diffusion (grey) calculated for 1 second duration and a 1 x 1 x $10^{-6}$ cm$^2$/sec D value. The "diffusion detection area", resented by the black and grey features covers 1.9 % of the total image area.

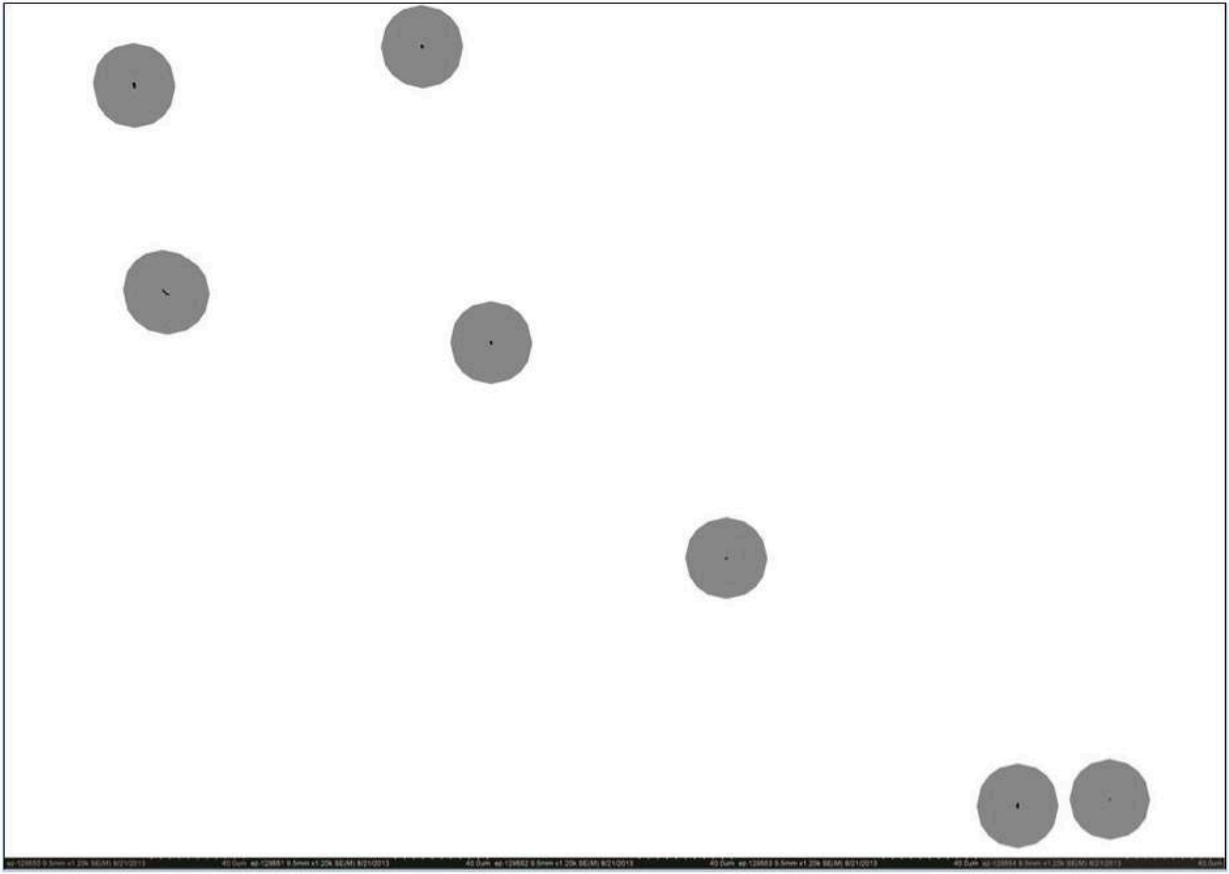

**Supplementary Fig. 10 | 2D-diffusion calculation for macroscopic defects II.** The resulting image after increasing the size of the pinholes (black) on all sides by the proton's radius of diffusion (calculated for a 1 second duration and a $1 \times 10^{-6}$ cm$^2$/sec D value) and the 3$\mu$m distance to account for the partial overlap of the laser spot with the "diffusion detection area" (grey). This resulting "diffusion detection/laser spot area", resented by the black and grey features, covers 2.9 % of the total image area.

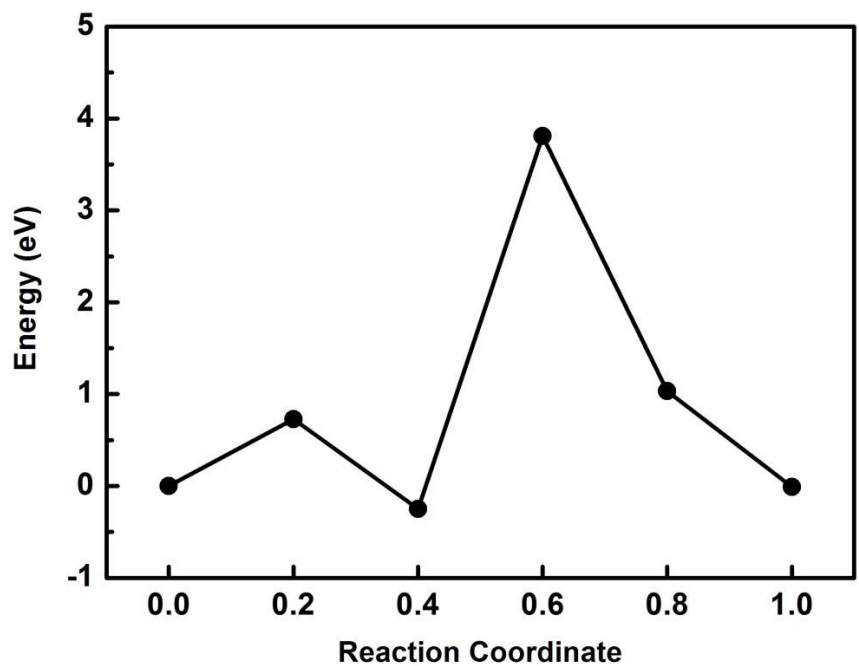
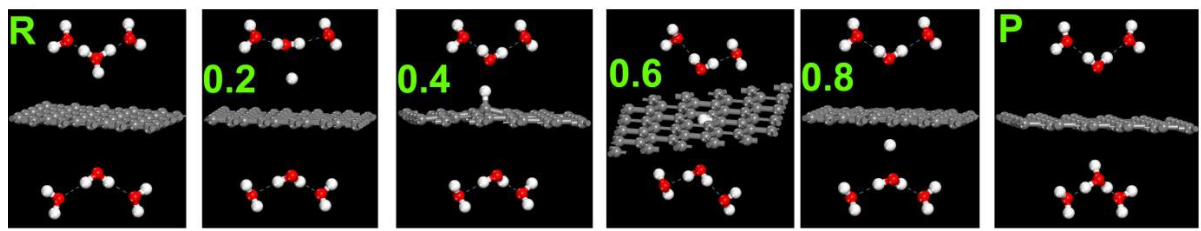

**Supplementary Fig. 11 | Aqueous Proton transfer through pristine graphene.** Energy barrier diagram and snapshots from nudged elastic band (NEB) calculations from reactant state "R" to product state "P" in fractional steps of 0.2.

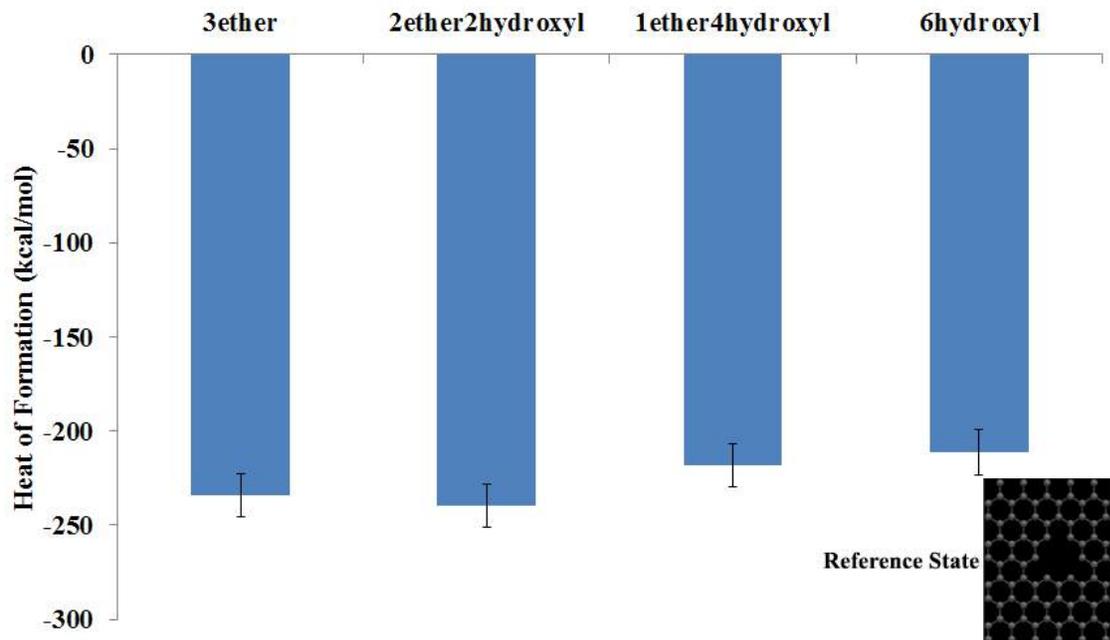

**Supplementary Fig. 12 | Stability of defect terminations.** Heats of formation for various defect terminations on quad-vacancy graphene.

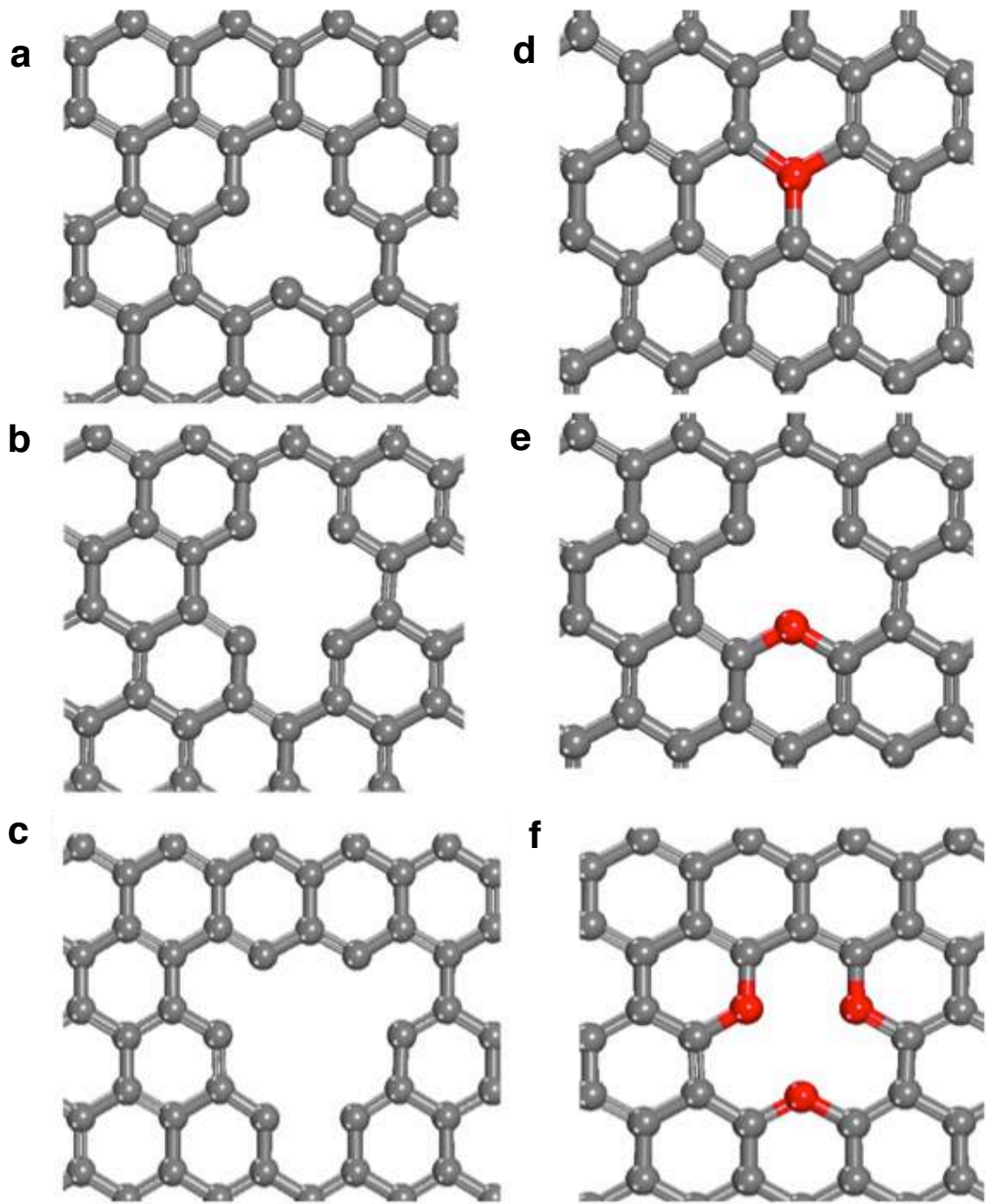

**Supplementary Fig. 13 | Atomistic views of some of the coordinatively unsaturated defect sites considered here.** The 1, 2, 4 (4V) carbon vacancy sites and the oxygen terminated vacancy sites in graphene.

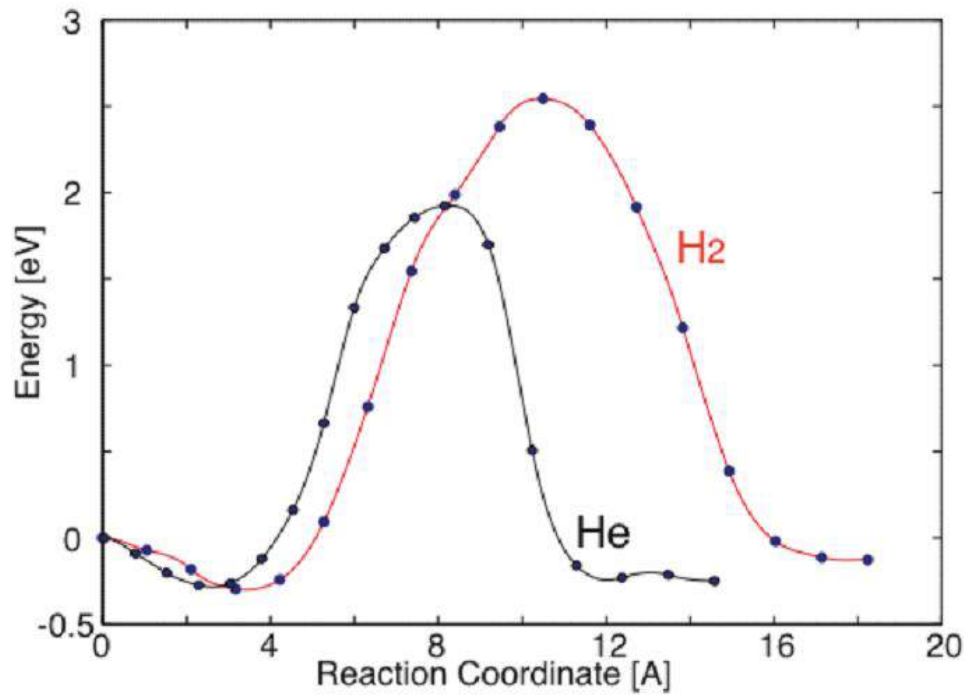

**Supplementary Fig. 14 | He and H$_2$ transfer across OH-terminated defect site.** DFT-calculated energetics for He and H$_2$ transfer through hydroxyl-decorated graphene indicate the barrier for He diffusion is 1.8 eV while that for H$_2$ is over 2.5 eV through the OH-terminated 4V site, which indicates that neither of these species would transfer through graphene at room temperature.

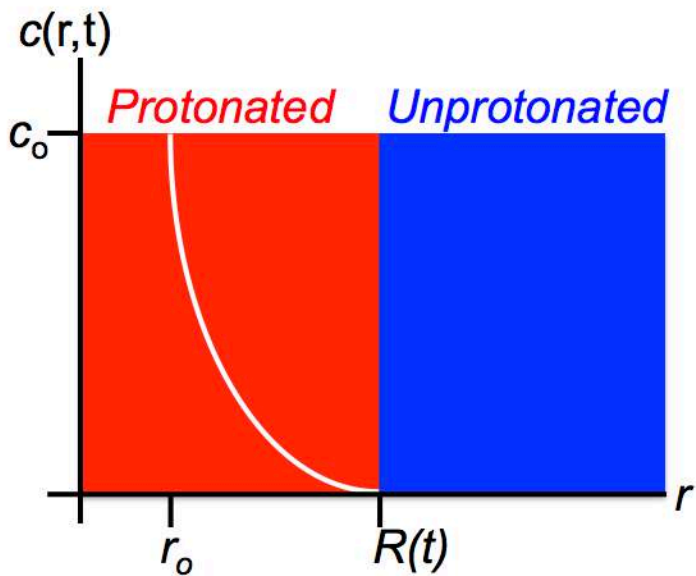

**Supplementary Fig. 15 | Diffusion calculation.** Sketch of the concentration vs distance profile and associated boundary conditions used for the estimation of the times required to propagate the protonation reaction front over a certain distance.

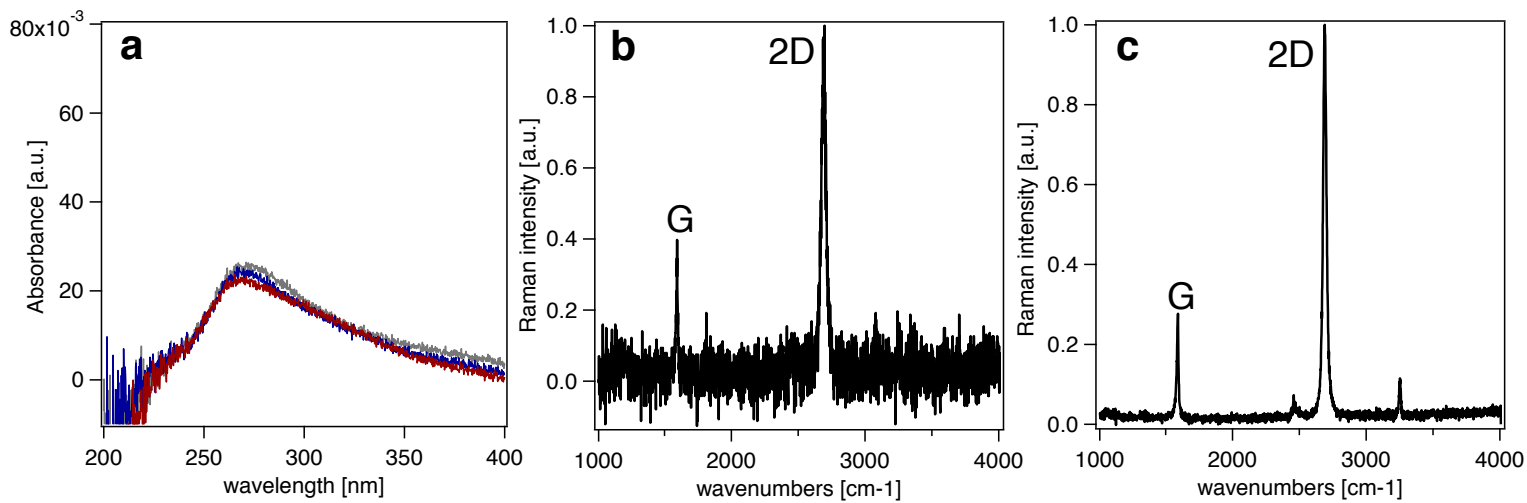

**Supplementary Fig. 16 | Stability assessment. (a)** Adsorption spectra of a single layer graphene film on a fused silica window before contact with aqueous solutions (grey trace), after a 20 minute soaking in a 1 mM NaCl Millipore water solution adjusted to pH 3 (red trace), and after a subsequent 20 minute soak adjusted to pH 11. Each trace is the average of 7 spectra collected over 7 different spots on the graphene film. Representative Raman spectra of a single layer graphene film **(b)** before SHG "pH jump experiments" and **(c)** after 2 days of SHG "pH jump experiments". The spectra are normalized to the highest intensity.

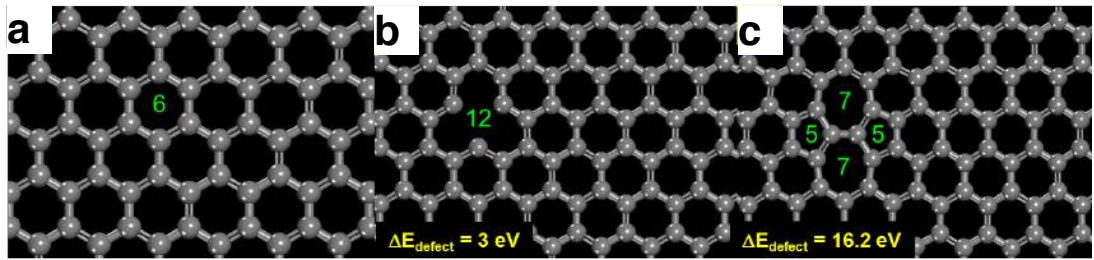

**Supplementary Fig. 17 | Defect energetics.** DFT-optimized structures for (**a**) pristine graphene, (**b**) 1V defect, and (**c**) Stone-Wales type defect sites and the cost to form the defect sites.

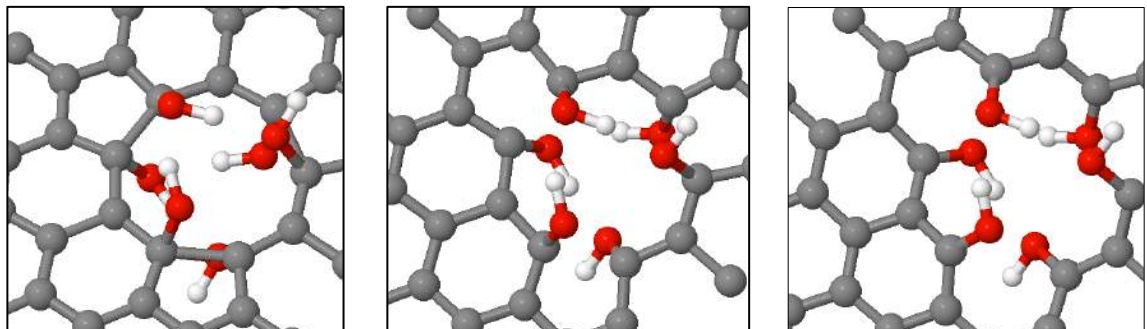

**Supplementary Fig. 18 | Relaxation of the defect sites.** (from **left** to **right**) The transition from the SW-type defect to the OH-4V-type defect upon hydroxylation: the SW defect was optimized first, followed by putting 6 hydroxyl groups near the defect with the -C-OH bond length being ~1.45 Å (near typical paraffinic C-O bond length). [Denotation: carbon in gray, oxygen in red and hydrogen in white]

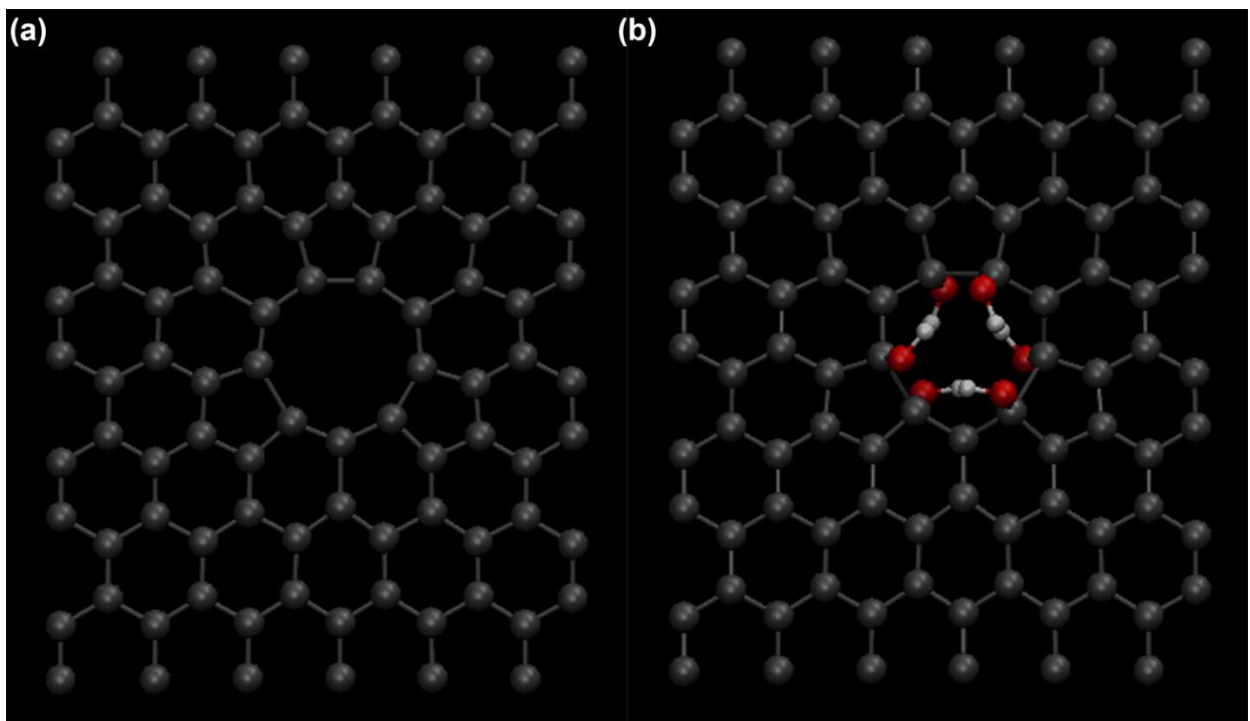

**Supplementary Fig. 19. | Defect relaxation and reconstruction. (a)** Reconstructed quad-vacancy defect in graphene and (**b**) reconstructed quad-vacancy defect terminated in 6 OH groups from ReaxFF calculations.

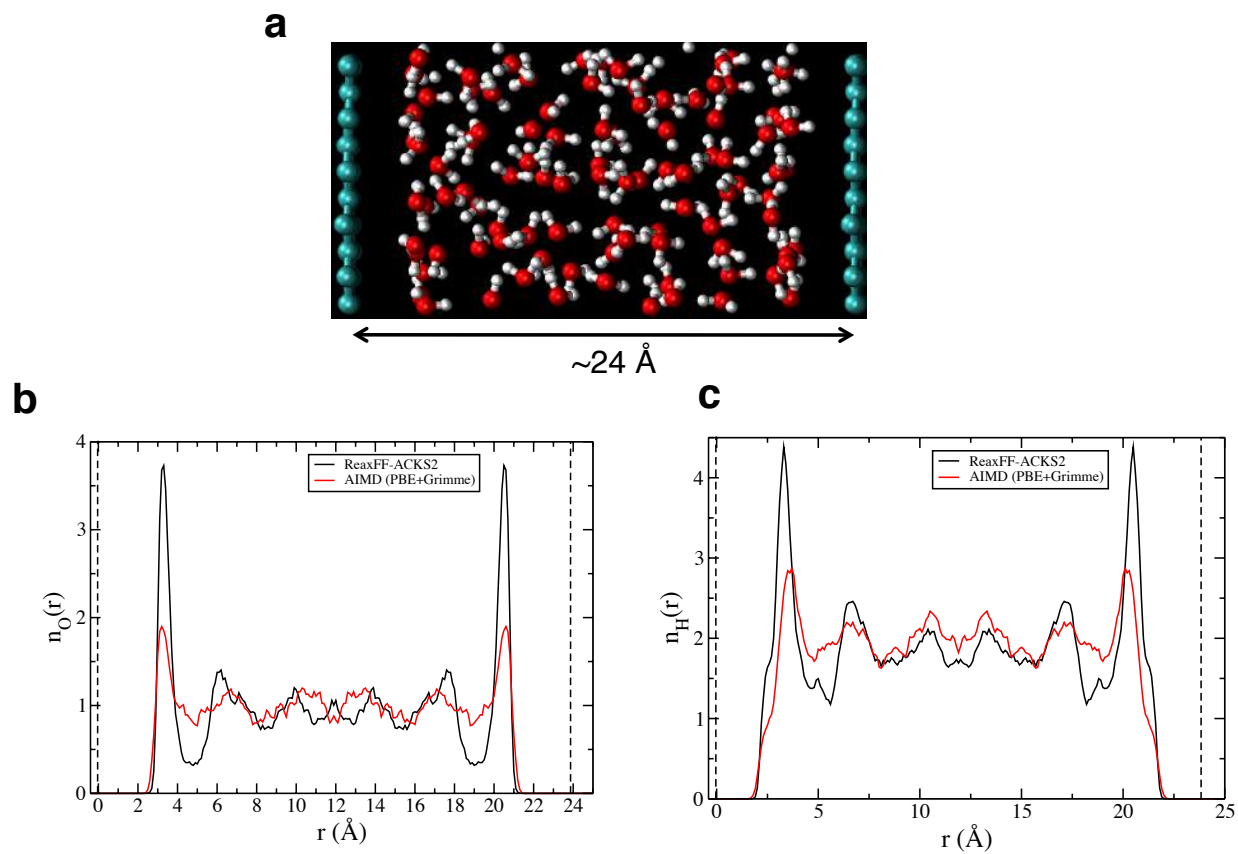

**Supplementary Fig. 20 | Structure of water confined between two graphene sheets.** (**a**) Graphene/water system used in the ReaxFF/AIMD simulations (**b**) oxygen number density profiles obtained from the ReaxFF/MD and AIMD simulations (**c**) hydrogen density profiles obtained from the ReaxFF/MD and AIMD simulations.

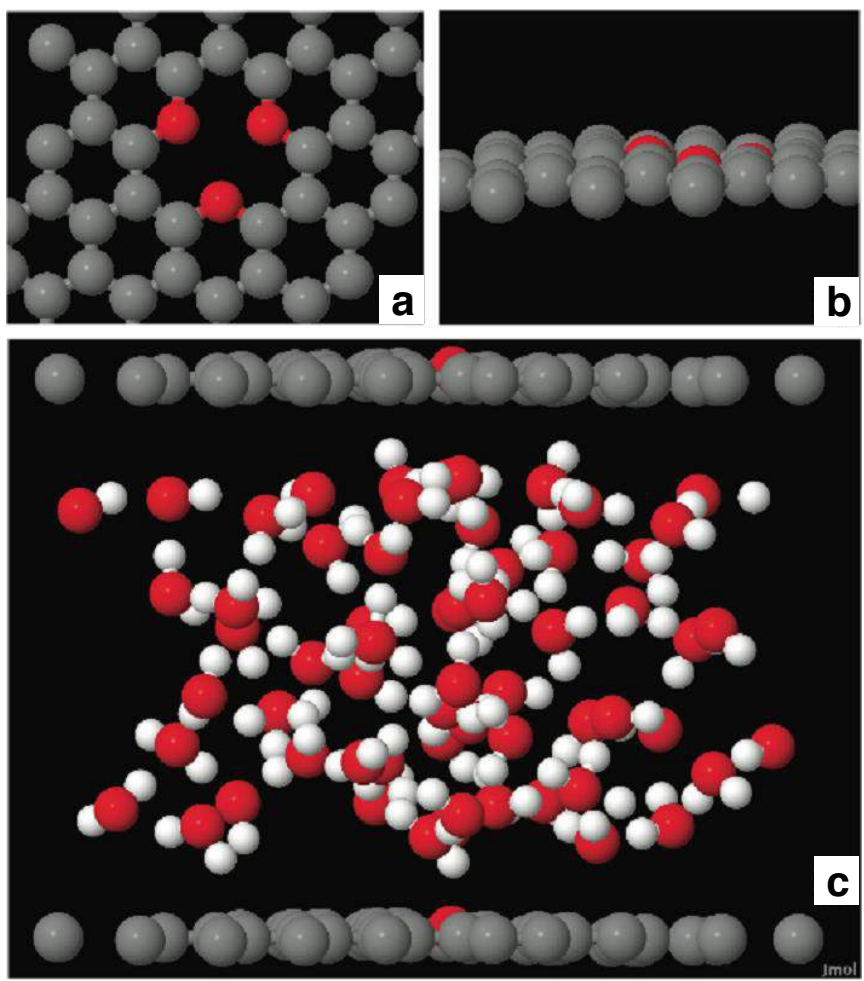

**Supplementary Fig. 21 | Ether-like pyrylium terminated defects.** The termination of the 4V-defect site in graphene with oxygen can result in the formation of 3 ether-like pyrylium sites. (**a**) top view; (**b**) side view; (**c**) at the water/graphene/water interface. The large grey spheres refer to carbon atoms whereas the, red, and white spheres refer to oxygen and hydrogen, respectively.

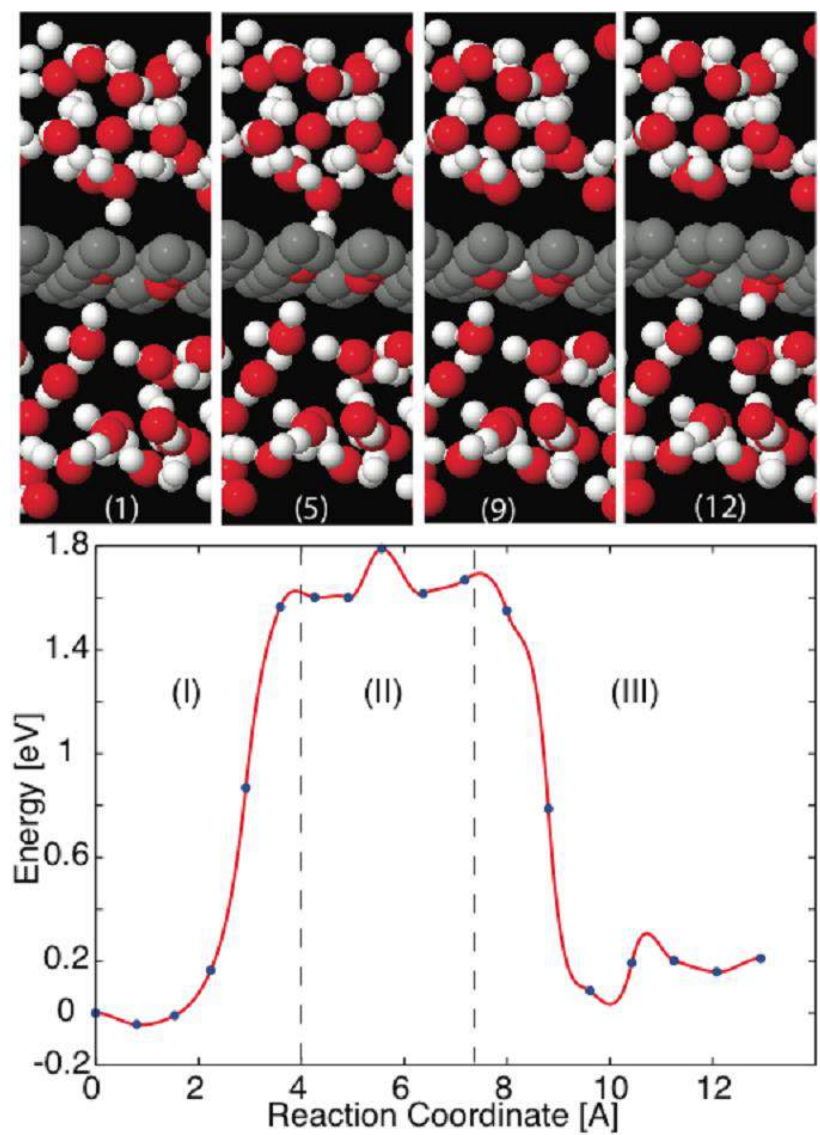

**Supplementary Fig. 22 | | Aqueous proton transfer through ether-like pyrylium terminated defects.** The reaction energy profile for proton transfer through the ether-decorated pyrylium sites in graphene: (region I) show 1) the release of proton from $H_3O^+$ to the pyrylium defect site in region II, and 2) relay of proton between ether groups; (region III) release of proton from ether to $H_3O^+$. [Snapshots from the NEB images are listed as insets on top with the image number marked at the bottom of each inset.]

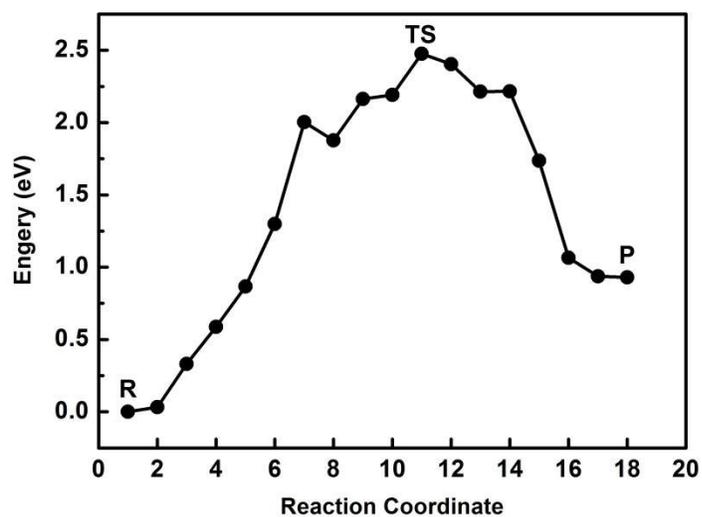

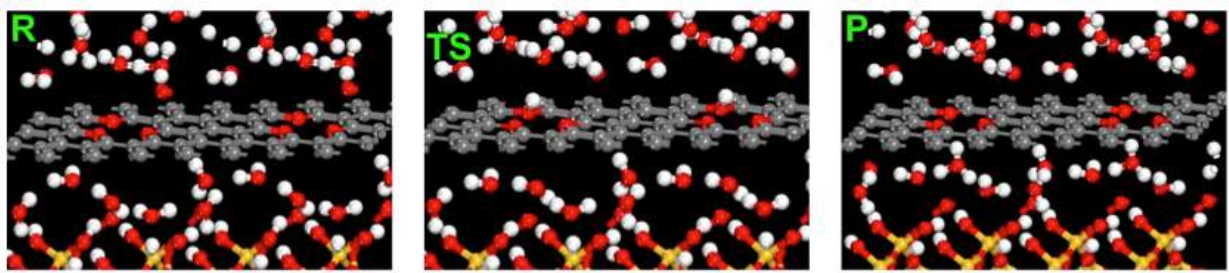

**Supplementary Fig. 23 | Proton diffusion through the 3O 4V pyrylium defect site that resides at the water/graphene/water/SiO$_2$ support interface.** The intrinsic barrier was calculated to be 1.8 eV whereas the barrier through the water/graphene/water interface was calculated to be 2.5 eV.

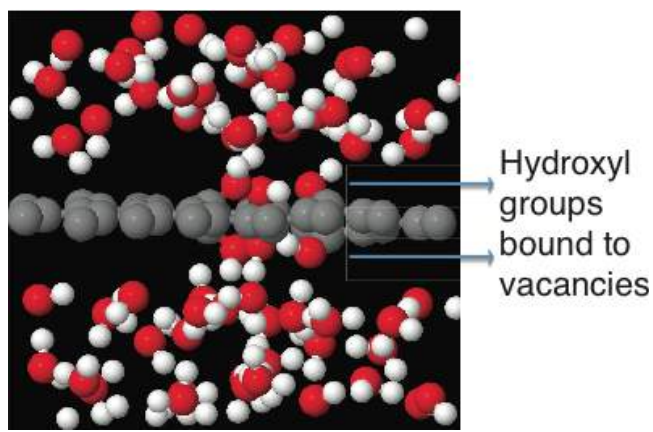

**Supplementary Fig. 24 | OH-terminated defect site.** The 4V vacancy defect site can react with water to functionalize the 6 coordinatively unsaturated carbon sites in the graphene basal plane with 6 OH groups. Three of the OH groups are oriented into the solution above the graphene surface while 3 are directed into the water solution below the graphene surface. This mixed OH/water interface provides an ideal conduit for proton transfer and proton shuttling.

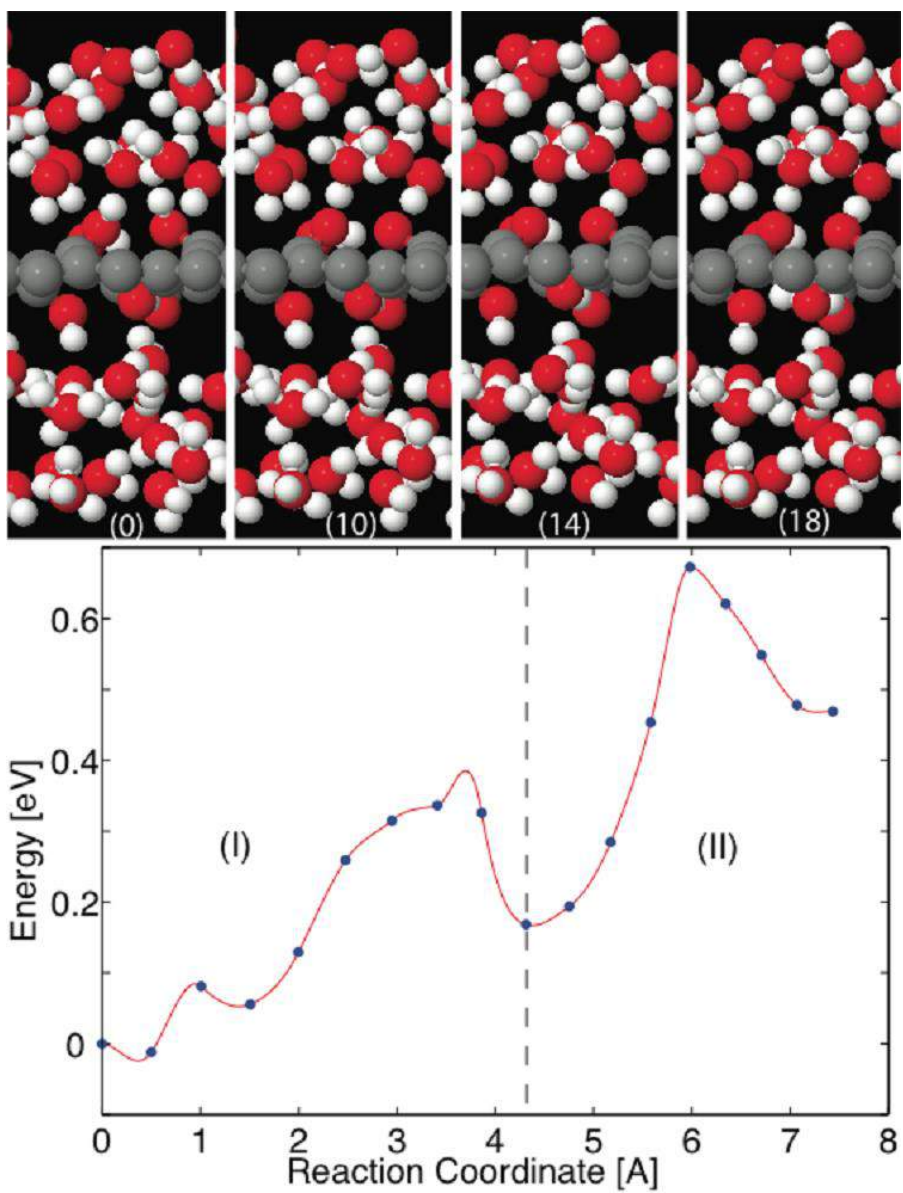

**Supplementary Fig. 25 | Aqueous proton transfer across OH-terminated defect site.** The energetics for proton transfer through the hydroxyl-decorated graphene: (region I) release of protons from $H_3O^+$ to hydroxyl/carbonyl groups; (region II) relay of proton between hydroxyl groups on different sides of the graphene sheet; (region III) <not shown here as it is the reverse processes in region I> release of proton from to hydroxyl/carbonyl groups to $H_3O^+$. Snapshots from the NEB images are listed as insets on top with the image number marked at the bottom of each inset.

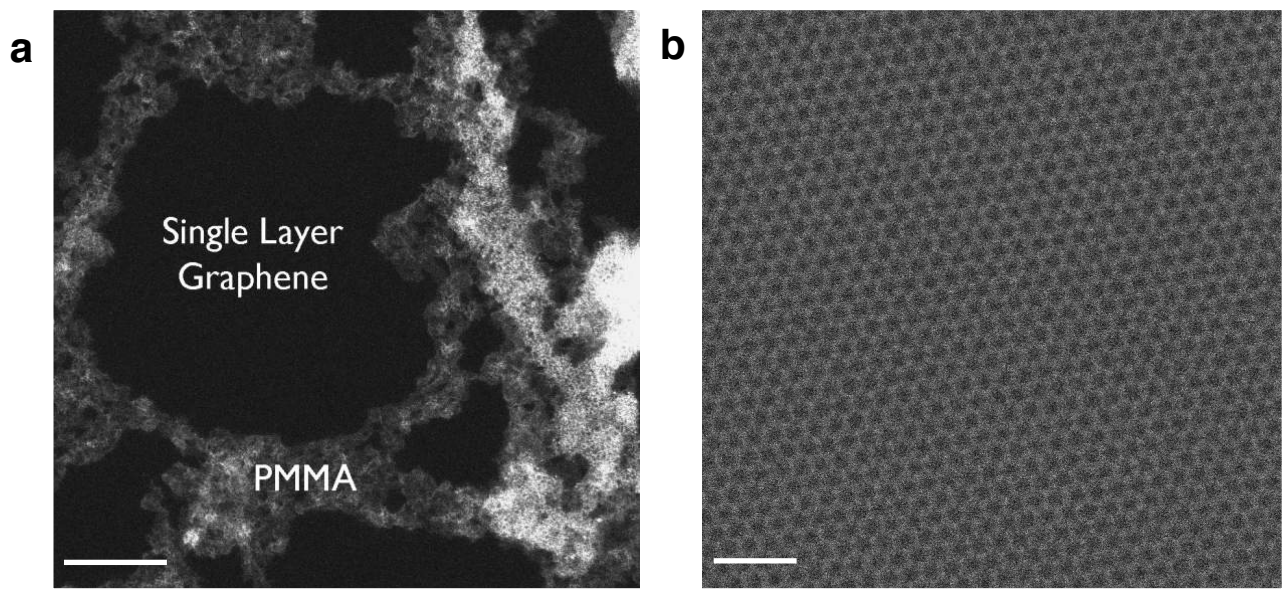

**Supplementary Fig. 26 | Imaging pristine graphene.** ADF STEM images of CVD prepared graphene showing (**a**) areas of single layer graphene between residual PMMA and (**b**) spatial resolved lattice structure of graphene within the labeled region. Scale bar in (**a**) is 10 nm, and in (**b**) it is 1 nm.

**Supplementary Table 1 | Duration of pH jumps as calculated from sigmoid fits.** The numbers in parentheses are the standard deviation of the durations.

|  | **Single layer graphene** | | **Fused silica** | |
| --- | --- | --- | --- | --- |
| pH jump | 3 → 10 | 10 → 3 | 3 → 10 | 10 → 3 |
| jump duration (sec) | 22(12) | 12(10) | 28(19) | 14(14) |

**Supplementary Table 2 | Rates of pH jumps as a function of flow rate.**

|  | Single layer graphene | | | | Bare fused silica | | | |
|---|---|---|---|---|---|---|---|---|
| pH jump | 3 → 10 | | 10 → 3 | | 3 → 10 | | 10 → 3 | |
| flow rate (mL/sec) | 0.9 | 0.3 | 0.9 | 0.3 | 0.9 | 0.3 | 0.9 | 0.3 |
| rate ($I_{SHG}$ counts/sec) | 2.5(1) | 2.6(4) | 1.6(1) | 0.7(2) | 3.7(1) | 2.0(3) | 1.8(1) | 0.7(2) |
| corrected rate ($I_{SHG}$ counts/sec)[a] | 1.8(1) | 2.7(4) | 1.2(1) | 0.8(3) | 2.6(7) | 2.1(3) | 1.3(1) | 0.7(2) |

[a] The resulting rate correcting for relative rate of analyte bulk concentration. Corrected rate values are obtained by correcting for the relative rate of changing ion bulk concentration as reported in ref [17] using the same experimental setup.

**Supplementary Table 3 | Calculated probability that the SHG response is due to pinholes.**

| Time (sec) | Probability |
|---|---|
| 1 | 3.5 % |
| 2 | 6.1 % |
| 3 | 8.2 % |
| 5 | 12 % |
| 10 | 21 % |

**Supplementary Notes**

**Supplementary Note 1: Analysis of the Magnitude of $I_{SHG}$ Change and the Time Duration of pH Jumps.**

As reported previously, SHG intensities are highly sensitive to changes of the interfacial potential, which is produced in the presence of surface charges[1-5]. The SHG response as a function of pH over charged silica surfaces has been studied previously, and it is well understood and accepted that the intensity of the SHG signal is directly related to the population of charged species on the silica surface[1,2,6-8]. Since the point of zero charge (PZC) of fused silica lies between a pH of ~ 2 and 3.5[7,9-16], as the solution pH increases from the PZC pH to pH 10, silanol groups will be deprotonated and the population of negatively charged groups at the surface will increase. In a report by Duval et al.[9], the relative surface population of three surface species, $SiOH_2^+$, $SiOH$, and $SiO^-$, was quantified using XPS for samples that had been exposed to solutions at pH values ranging between 0 and 10. More specifically, Duval reported silanol group population densities at pH 3 of approximately 7% for $SiOH_2^+$, 83% for $SiOH$, and 10% for $SiO^-$ and population densities at pH 10 of approximately 5% for $SiOH_2^+$, 70% for $SiOH$, and 25% for $SiO^-$[9]. Thus, with an increase in the solution pH, the SHG signal will increase due to a higher interfacial potential induced from the larger negative surface charge density.

As discussed earlier, the SHG signal in the pH jump experiments is collected as a function of time for a given solution pH value with a constant 1mM NaCl background over both a bare fused silica and a pristine single layer graphene sheet transferred onto a fused silica window. As shown in Fig. 1b in the main text, it is evident that the SHG response of the two systems is very similar. We quantified this similarity by first comparing the relative SHG intensities, $I_{SHG}$, obtained for the pH jumps for the two systems. To compare the magnitude of $I_{SHG}$ for the two systems, we normalized the SHG E-fields, given by $E_{SHG} = \sqrt{I_{SHG}}$, to the average $E_{SHG}$ at pH 7 from the beginning and end of each pH jump experiment. The averaged SHG E-fields at pH 3, pH 7, and pH 10 for the single layer graphene system and the bare fused silica system are shown in bar graph format in Supplementary Fig.1. This bar graph is comprised of the compiled data for 3 experiments on 2 different graphene samples (blue), and 4 experiments on 2 different silica samples (red). Supplementary Fig. 1 shows that the magnitudes of the normalized SHG E-fields at the three pH values are within error of one another. This similarity indicates that the populations of the charged surface groups for both the graphene and bare fused silica systems are similar.

For our second analysis we examined the duration of the pH jumps between the two systems. As shown in Supplementary Fig. 2 the pH jumps from 3 to 10, and likewise from 10 to 3 both the graphene (blue trace) and bare fused silica (red trace) appear to reach completion within the same time span. In order to quantify the duration of the jumps in both systems, we calculated the time it took for the SHG intensity to increase from 10% to 90%, which was accomplished by referencing each jump to the average SHG intensity at pH 3, and then normalizing each jump to the average SHG intensity at pH 10. Using IgorPro software, the $I_{SHG}$ vs. time trace was then fit with the following sigmoid:

$$y = b + \frac{n}{1+e^{\frac{(\tau-x)}{r}}} \qquad (1)$$

Here, $y$ is the normalized SHG intensity, $b$ is a fit parameter related to the initial $I_{SHG}$ value, $n$ is a fit parameter related to the difference between the initial and final $I_{SHG}$ values, $r$ is the rate by which $I_{SHG}$ changes in time during the jump, $\tau$ is the inflection point, and $t$ is time in seconds. We then solved for the corresponding time value at the 0.1 and 0.9 values of the normalized SHG intensity, and averaged the time durations for the 3 to 10, and 10 to 3 jumps respectively for both the graphene and bare fused silica systems. The results for this analysis are summarized in Supplementary Table 1. Considering the time durations calculated here, it is evident that there is no appreciable delay between the graphene and silica systems within the sensitivity of our system.

For our third analysis we compared the rates from the sigmoidal fits for pH jump experiments conducted with flow rates of 0.9 mL s$^{-1}$ and 0.3 mL s$^{-1}$. The rates for the two pH jumps with both flow rates are summarized in Supplementary Table 2. Following the method described in a previous publication[17], we also reported the rates corrected for the concentration profile of the protons/hydroxides as a function of time. The rates for the pH jumps do not differ between the single layer graphene and bare fused silica systems for either the slow or fast flow rates. Additionally, these results show little appreciable dependence of the rates on the flow rate, which will be important for the diffusion kinetics analysis discussed below. Based on these three analyses we conclude that the two systems do not differ in terms of relative surface charge density, in the duration of the jumps, or in the rates of the jumps, which indicates that the acid/base chemistry at the fused silica interface occurs in an unimpeded fashion in the presence of graphene.

**Supplementary Note 2: Determination of pKa Values by SHG.**

To rule out the possibility that the $I_{SHG}$ response over the graphene film is due to proton/hydroxide adsorption above the graphene sheet, the SHG technique was applied to test for the presence of the two well-known acid-base equilibria that have been reported in the literature for fused silica/water interfaces[1,16,18,19]. Specifically, fused silica surfaces contain two types of silanol sites having pKa values ranging between approximately 3.76 and 4.5 for the more acidic sites, and 8.3 and 10.8 for the less acidic sites, depending on the electrolyte identity and concentration[1,16,20]. The existence of these two silanol groups is attributed to two populations of silanol groups involved in either weak or strong hydrogen bond formations with interfacial water molecules[16,20-22]. Experimental evidence for bimodal acid-base equilibria can confirm that the observed SHG response with jumps in bulk pH is due to the protonation and deprotonation of silica surface sites.

The SHG pKa experiments were run on the same experimental setup as described for the SHG pH jump experiments, except that all aqueous solutions contained a 100 mM NaCl background electrolyte solution. The SHG intensity was then collected as a function of pH starting from pH 11 and ending at pH 2.75. As with the pH jump experiments, the desired pH was adjusted and maintained throughout the duration of the experiment with minimum amounts of dilute solutions of ~1M NaOH and HCl. For each pH evaluated, the SHG intensity was allowed to reach steady state at which point the SHG signal was collected for a duration of at minimum 300 seconds. The resulting normalized SHG E-field is plotted as a function of solution pH in Supplementary Fig. 3A. The averaged E-field for each pH is normalized to the averaged E-field at pH 7. In Supplementary Fig. 3A, the resulting $E_{SHG}$ values were then referenced to the minimum SHG E-field around pH 3 and then normalized to the maximum SHG E-field around pH 11. The SHG E-field increases as the bulk pH increases, except that there is a plateau of the SHG E-field starting around pH 6 before another sharp rise in the SHG E-field around pH 8. This plateau and the two inflection points around pH 5 and 9 are consistent with the existence of two surface sites and two pKa values.

Using the effective pKa, ($pK_a^{eff}$), calculations proposed by Azam *et al*. we can roughly approximate the surface pKa values correcting for the background electrolyte concentration. Here the $pK_a^{eff}$ is given by

$$pK_a^{eff} = p(K_a K_{assc}) = pH_{0.5} - pM^+ \qquad (2)$$

where $K_a$ is the acid dissociation constant, $K_{assc}$ is the background electrolyte cation ($M^+$) association constant, $pH_{0.5}$ is the pH where the normalize SHG E-field is equal to 0.5 for a given type of silanol, and $pM^+$ is given by $-\log[M^+]$. In Supplementary Fig. 3B and 3C the normalized SHG E-field is plotted as a function of pH for the more acidic and less acidic silanol groups, respectively. Fitting these traces with a sigmoid curve, the inflection points at $pH_{0.5}$ were used to calculate the $pK_a^{eff}$ for the single layer graphene system in the presence of 100 mM NaCl. We calculated $pKa^{eff}$ values of 3.5(1) and 8.3(2), which fall within the reported literature values[16]. The experimental validation of the existence of two silanol sites and their approximate pKa values confirm that the SHG response is due to acid-base chemistry occurring at the silica surface.

**Supplementary Note 3: SHG pH Jump Experiment on Multilayer Graphene.**

A multilayer graphene film was prepared by growing bilayer graphene films, and then transferring the bilayer graphene films on top of one another to a silica window several times until an approximate 8-layer graphene film was formed. The normalized Raman spectrum for the multilayer graphene film is shown in Supplementary Fig. 4. The Raman spectrum of the multilayer graphene film exhibits the G and 2D bands associated with graphene films, and a small D band at 1350cm$^{-1}$ indicative of defects[23,24]. Raman spectra were collected over four different spots on the graphene film and exhibited an average G:2D band ratio of 3.1(7) and 2D band full width half max of 86(1) cm$^{-1}$, where the number in parentheses is the standard error on the point estimate. These values are consistent with multilayer graphene[23-30]. We did not attempt to exactly quantify the number of layers in the graphene film as Raman spectroscopy has been reported to only accurately identify the number of graphene sheets between 1 and 5[23].

The SHG pH jump experiment was then carried on the multilayer graphene. A plot of the SHG intensity as a function pH and time is shown in Supplementary Fig. 5. As with the single layer graphene pH jump results, we see that there is no delay or attenuation of the SHG response with the multilayer graphene (black trace) when compared to the bare fused silica (crimson trace) and the single layer graphene (blue trace). Since the multilayer graphene film was prepared by sequential deposition of ~ double layer graphene films we propose that the quick diffusion process through the multilayer film may be due to proton diffusion though a brick-and-mortar-like structure of the multilayer graphene film, similar to recent work on unimpeded water permeation through graphene oxide films[31]. A brick-and-mortar structure could be formed due to the likelihood of grain boundaries present in the graphene film and the possible water layers that get trapped between the graphene layers during the transfer process.

**Supplementary Note 4: Proton Diffusion.**

For complex systems with multiple potential proton transfer sites, ReaxFF provides a suitable alternative to the multistate empirical valence bond (MS-EVB) method developed by Voth and coworkers that can successfully describe the proton transport in water and biomolecular systems[32]. For an extremely complicated system with multiple potential reactions such as that under consideration here, the MS-EVB is no longer applicable because this approach requires a combination of states, which represent the chemical bonding topologies involving the proton. In our applications, we have used the ReaxFF method for water/proton/silica systems[33]. In addition, ReaxFF also reproduces experimental results for both water self-diffusion and proton diffusion in water[34]. We therefore employed ReaxFF MD simulations to calculate proton and water self-diffusion constants in water confined between graphene and α-quartz (001) surface. The diffusion constants provide an estimate of the area accessible via proton diffusion during a time $t$ from which we compute the probability that the rise and drop of the SHG response over graphene is due to pinholes.

ReaxFF, being a reactive force field, takes chemical reactivity into account, which, as discussed below, is of significance in describing proton diffusion in the presence of surface SiO⁻ species. The self-diffusion coefficient of water at 298K given by ReaxFF ($2.1 \times 10^{-5}$ cm$^2$ s$^{-1}$) is comparable to the values given by diaphragm-cell technique[35] ($2.272 \times 10^{-5}$ cm$^2$ s$^{-1}$) and pulsed magnetic field gradient (PFG) NMR ($2.299 \times 10^{-5}$ cm$^2$ s$^{-1}$) studies[36]. ReaxFF also reproduces the proton diffusion constant ($9.04 \times 10^{-5}$ cm$^2$ s$^{-1}$) in water at 298 K, which is in good agreement with experimental studies[37] ($9.31 \times 10^{-5}$ cm$^2$ s$^{-1}$). DFT and ReaxFF calculations predict water dissociation on a 1V site in graphene to be exothermic with an activation barrier of 47.79 kcal/mol and 36.83 kcal/mol respectively.

The simulated system consists of water molecules sandwiched between graphene and quartz surfaces separated by 1.4 nm as shown in Supplementary Fig. 6. In our simulations we used a periodic quartz (001) slab with (6 x 4) unit cells parallel to the surface with 2 layers of SiO$_2$ and a periodic graphene sheet with (10 x 5) unit cells parallel to the surface. Water molecules and protons were placed in random configurations between the quartz and graphene surfaces. The dimensions of the simulation cell are 24.73 Å x 13.84 Å parallel to the surface and 25 Å in the direction perpendicular to the surface.

The system was energy minimized with convergence criterion of 0.5 kcal/Å and equilibrated in in the canonical (NPT) ensemble for 100 ps at 300K, with a time step of 0.25 fs using the Berendsen thermostat with a coupling time constant of 100 fs and Berendsen barostat with a coupling time constant of 500 fs to control temperature and pressure of the entire system. We perform MD simulations on the equilibrium configuration for 125 ps at 300K to calculate the proton and water self-diffusion constants. In these simulations the equilibration time was 25 ps and the subsequent production run was 100 ps.

Since the Grotthuss mechanism plays a key role in the proton transport in solution, we evaluated the sequence of proton transport events in our system[38]. To do so, the oxygen ($O^*$) of the water with an excess proton is distinguished from the neutral water molecules. The same index number of $O^*$ between two adjacent frames from MD trajectory indicates the vehicular transport without proton hopping. A change in the index of the $O^*$ between frames indicates occurrence of Grotthuss-type hopping. Using a time-dependent trajectory of $O^*$, the diffusion constant of proton transport is calculated based on mean-square displacement and the Einstein relation.

We consider partially hydroxylated (40% SiOH) and fully hydroxylated (100% SiOH) α-quartz(001) surfaces for calculation of the diffusion constants. The self-diffusivity of nano-confined water is reduced roughly by a factor of 24 for the partially hydroxylated (9.537 x $10^{-7}$ $cm^2 s^{-1}$) case and is comparable to liquid water for the fully hydroxylated (2.892 x $10^{-5}$ $cm^2 s^{-1}$) case. Supplementary Fig. 6 gives the density plot of oxygen (blue) and hydrogen (red) atoms in the simulation box during a NVT run at 300 K for (a) partially-hydroxylated and (b) fully-hydroxylated case. As shown in Supplementary Fig. 6, the partially hydroxylated case shows a strongly enhanced 'ice-like' local water structure close to the silica surface whereas the fully hydroxylated case shows a more diffuse water structure similar to liquid water. The O-O separation (2.47 Å) for SiO⁻···HOH hydrogen bonds at the partially hydroxylated quartz surface is significantly shorter than O-O separation in bulk water (2.83 Å) indicating the presence of strong hydrogen bonds at the partially hydroxylated quartz/water interface. These strong hydrogen bonds give rise to the 'ice-like' water structure at the partially-hydroxylated quartz/water interface and leads to very low water self-diffusion near the silica surface. The silanol groups on the fully hydroxylated surface form intralayer hydrogen bonds with surface O atoms on the adjacent row in the V-shaped ridges and do not interact with the water phase above, giving rise to a vacuum gap between the fully hydroxylated quartz surface and the water phase as shown in the right panel of Supplementary Fig. 6. This finding explains the absence of a local water structure and faster self-diffusivity at the fully hydroxylated quartz/water interface.

In simulations containing partially hydroxylated quartz surfaces the proton diffusion is quickly terminated by protonation of the surface SiO⁻ groups. This result indicates that proton diffusion is significantly slower in the presence of surface anionic species due to proton trapping at these sites and the slower water self-diffusion (~factor 24 reduction compared to liquid water). XPS studies by Duval *et al*.[9] report that surface SiO⁻ species are present on quartz in the pH range 1 to 10, indicating that proton diffusion is significantly lowered for the entire studied experimental pH range. For the fully hydroxylated case the proton diffusion constant is reduced roughly by a factor of 2 compared to bulk water ($4.944 \times 10^{-5}$ cm$^2$/s).

**Supplementary Note 5: SEM Pinhole and Proton Diffusion Analysis.**

When considering the pH jumps, it could be possible that the focused 30 $\mu$m laser spot is centered directly over a pinhole. In this scenario, the presence of a pinhole could account for the similarity of the SHG response between the bare silica and graphene systems. In this case, the protons could easily pass directly though the pinhole and access the silica surface sites in order to participate in the acid-base chemistry needed to account for the rise and drop in the SHG signal intensities. The probability that our SHG response originates from a pinhole allowing for proton transfer from above the graphene film to the silica surface underneath can be calculated by determining the probability that our SHG beam is focused over a pinhole. However, the probability we need is not simply the percent area of the SEM image covered by pinholes, instead, the area accessible via proton diffusion at time $t$ and the area of the focused laser beam also need to be taken into consideration.

For our analysis, SEM images were collected on the graphene film to detect pinholes. The SEM images were collected in a grid-like manner in order to obtain a continuous 529 $\mu$m x 397 $\mu$m area image at the center of the graphene film. To determine the pinhole density, the images were combined as shown in Supplementary Fig. 7, which is representative of a central area of the graphene film where the SHG experiments are conducted. In order to compensate for the difficulty in identifying pinholes due to charging effects, the image was analyzed by eye, and any anomaly observed using ImageJ software that was greater than 9 pixels in size was outlined, and then marked as a pinhole by setting the outlined feature to be white (white value=255, 255, 255). For some of these anomalies it was not clear if the imaged feature was a pinhole or a contaminant, such as a piece of dust, debris, or detritus. However, it was decided to treat these features as pinholes and thus compute an upper bound of the pinhole density. Any anomalies smaller than 9 pixels in size were not counted as pinholes as these features are indistinguishable from the image noise. Following thresholding of the brightness to values larger than 254 in ImageJ, we quantified the area that the pinholes covered with the "Analyze Particles" feature in said software. Seven anomalies were detected covering an area of 14 $\mu m^2$, or 0.007% of the total image area, as shown in the returned ImageJ "masks" image in Supplementary Fig. 8.

The probability of placing our laser beam within the "diffusion area" emanating from a macroscopic pinhole was then calculated for arbitrarily chosen times of 1 second and 10 seconds by effectively increasing the size of the pinholes on all sides by the proton's radius of diffusion. The pinholes sizes are graphically increased using Adobe Photoshop while transforming the ImageJ scale bar to the pixel equivalent for Photoshop (5.7 $\mu$m pixel$^{-1}$). ImageJ was then used to determine the total area covered by the added pinhole size, which represents the "diffusion detection area", or the area of the film, which, if probed, could account for the observed SHG response. An example of this alteration is shown in Supplementary Fig. 9 for 1 second duration and a 1 x 10$^{-6}$ cm$^2$ s$^{-1}$ D value. In this case, the probability that our SHG response is due to proton exchange through pinholes and subsequent diffusion from there to the laser spot is 2.3 %. Lastly, it is also necessary to take into consideration that the laser beam is not completely overlapped with the "diffusion detection area". In that case, the SHG response may not be significantly altered if diffusion were indeed to be the operative mechanism. The largest standard deviation of the SHG signal intensity is 15%, and therefore, an overlap of less than 85% should result in a distinct observable change in the SHG signal intensity, adding approximately 2 to 3 $\mu$m of the total diameter of the laser spot to the "diffusion detection area". We therefore increased the size of the pinholes on all sides by the distance of the proton diffusion radius plus the 3 $\mu$m distance to account for partial overlap of the "diffusion detection area" and the laser spot. Again, using ImageJ, we calculated the percent area covered by the new "diffusion detection/laser spot area". The resulting image for this analysis is shown in Supplementary Fig.

10 (1 second duration, $D = 1 \times 10^{-6}$ $cm^2$ $s^{-1}$). In this case, the final probability that our SHG is due to pinholes is 3.5 %.

The "masks" image shown in Supplementary Fig. 8 was also used to quantify the average pinhole size and the average distance between pinholes. Using the ImageJ software, the "Analyze Particles" function was used to count and output particle area along with the x and y coordinates of the center of mass for each detected particle. Appropriately accounting for the dimensions of the image and the SEM image scale bar, the average area and distance between pinholes was calculated for the 7 detected pinholes. The average area of the pinholes was calculated to be 2(1) $\mu m^2$. The distances from each pinhole to every other pinhole was calculated based on the x and y coordinates provided in the "Results" output of the "Analyze Particles" program. These distances were then averaged together to provide the reported pinhole-to-pinhole distance of 300(160) $\mu m$. Here, the number in the parenthesis is the standard error on the point estimate.

**Supplementary Note 6: ReaxFF Calculations of Heats of Formation for the Considered Ether/Hydroxyl Defect Terminations.**

We performed MD-NVT simulations for the various ether/hydroxyl group terminations (3ether, 2ether +2OH, 1ether+4OH, 6OH) considered in this study in the presence of a bulk aqueous phase (96 water molecules) to investigate their relative stabilities at room temperature. We consider a periodic bare quad-vacancy graphene sheet with (10 x 5) unit cells parallel to the surface in contact with a bulk aqueous phase containing 96 water molecules as the reference state. The bare quad-vacancy graphene/water system is considered as the reference state since the graphene sheet utilized in the experimental studies contains defect sites. The various defect terminations were manually prepared by introducing them at the defect site and solvating the excess protons in the aqueous phase. The systems were energy minimized with a convergence criterion of 0.5 kcal/Å and equilibrated in the canonical (NVT) ensemble for 100 ps at 298.15K, with a time step of 0.25 fs using the Berendsen thermostat with a coupling time constant of 100 fs. The total system energy averaged over the final 20 ps of the NVT simulation was used to calculate the heats of formation. Supplementary Fig. 12 gives the heats of formation for the various defect terminations, obtained by considering the bare quad-vacancy graphene/water system as the reference state. The 2ether+2OH termination is the most stable state followed by 3ether, 1ether+4OH and 6OH terminations. All the considered defect terminations are energetically favorable as compared to the bare quad-vacancy system. We note that energy barriers to the formation of the defect sites were not computed.

**Supplementary Note 7: Helium and $H_2$ Diffusion Through Graphene.**

Both DFT and ReaxFF simulations indicate that the diffusion of He and $H_2$ is quite difficult and as such the 4V vacancy sites are essentially exclusive for proton transfer. DFT-calculated activation barriers for He diffusion through the O- and OH- defect sites were found to be greater than 1.8 eV. The larger size of He and its inability to hydrogen-bond prevent it from diffusing through these atomic scale vacancies. The results clearly show that He cannot form hydrogen bonds and thus requires much higher temperatures to overcome activation energies that are > 2.0 eV to diffuse through the 4V sites (see Supplementary Fig. 14 black trace). The results are also consistent with experimental observations and help to rationalize the proton transfer mechanisms, kinetics and sites of interest in graphene.

Our calculations indicate that the DFT-calculated barrier for $H_2$ to diffuse through the hydroxyl-decorated vacancies sites for the 4V defect of over 3.0 eV (see Supplementary Figure 14 red trace). This is consistent with the results from He and the fact that the larger size of $H_2$ causes more repulsion, thus resulting in a much higher barrier. Our studies suggest that the proton is the only hydrogen species that can diffuse through the graphene sheet. The $H_2$ molecule is too large and cannot take advantage of hydrogen bonding.

The results on the size and defect termination requirements for proton, He, and $H_2$ transfer provide important insights into the properties of graphene and suggest that single layer graphene may be used as a membrane by which to selectively separate protons from a wide range of other species and as such may be useful in various electrochemical processes such as in fuel cells or batteries without any crossover.

**Supplementary Note 8: Estimates Regarding the Propagating Reaction Front.**

Diffusive transport of protons between the graphene and silica from an atomic defect of radius $r_0$ can be modeled with a pseudo steady state approximation where the propagation of the reaction front has negligible effect on the proton concentration distribution.[39] We apply radial symmetry and cylindrical coordinates to the general proton concentration, c, vs. distance, r, profile shown in Supplementary Fig. 15. Given the steady-state approximation, we neglect the time derivative in the general diffusion equation. We assume that the instantaneous concentration profile is the steady state solution for the following boundary conditions: $c=c_o$ at $r=r_o$, where $c_o$ is the proton concentration per unit area at the defect site, which has a radius $r_o$, and $c=0$ at $r=R(t)$. We then arrive at the following expression: $c(r, t) = c_o \cdot [\ln(r_o/R(t))]^{-1} \cdot [\ln(r) - \ln(R(t))]$.

For a diffusion coefficient, D, of protons moving between the graphene and silica, and a given number density of sites that can be protonated, $\gamma$, the rate of the propagating protonation front at radius R is given by $dR \cdot dt^{-1} = D \cdot c_o \cdot [\gamma \cdot R \cdot \ln(R/r_o)]^{-1}$. Solving for t, one arrives at $t = \gamma \cdot (D \cdot c_o)^{-1} \cdot [R^2 4^{-1} \cdot (2 \ln(R \cdot r_o^{-1}) - 1) + r_o^2 4^{-1}]$. For a given atomic defect site radius, $r_o$, of ~ 0.1 nm, a diffusion coefficient, D, of ~ $1 \times 10^{-10}$ m$^2$ s$^{-1}$, a site density of $\gamma \sim 1$ nm$^{-2}$, and a proton concentration at the defect site, $c_o$, of ~ $6 \times 10^{14}$ sites m$^{-2}$, one computes an estimated duration of ~ 1 s for filling an area having a radius R of ~ 100 nm. Slightly larger areas (R=300 nm) take ~ ten seconds to fill. While we caution that these estimates depend on the input values for $\gamma$, D, $c_o$, and r, the results seem reasonable to within a factor of two or three, given the approximations.

**Supplementary Note 9: Graphene Characterization and Analysis with Raman and UV-Vis Spectroscopy.**

Like previously reported[4,5], Raman and UV-Vis spectroscopies were used as methods to confirm that the graphene films were not altered due to the high and low pH conditions or due to the experimental procedures outlined above. For analysis by UV-Vis spectroscopy, a spectrum of the graphene film was collected prior to solution exposure (Supplementary Fig. 16a, grey trace), after soaking in a 1mM NaCl Millipore water solution at pH 3 for 20 minutes without rinsing (Supplementary Fig. 16a, red trace), and finally after soaking in a 1mM NaCl Millipore water solution at pH 11 for 20 minutes without rinsing (Supplementary Fig. 16a, blue trace). There is no apparent change between the three absorption spectra, confirming that the graphene film can withstand the high and low pH conditions.

For the Raman analysis, spectra were recorded with an Acton TriVista CRS Confocal Raman System using a 514.5 nm excitation wavelength with a 100x objective at a power density < $10^6$

W cm$^{-2}$ to avoid sample damage. Raman spectra were collected prior to SHG pH jump experiments and also after 5 consecutive days of SHG experiments, including 2 days of SHG pH jump experiments. The resulting normalized Raman spectra prior and post experiments are shown in Supplementary Fig. 16b and 16c, respectively. An in-depth analysis of the Raman spectra collected from the graphene films prepared for our SHG experiments is available in our previous publications[4,5]. The Raman spectra obtained from our samples are remarkably similar to those recently reported for epitaxially prepared wafer-scale graphene[40]. Moreover, the Raman spectrum collected from samples following pH jumps does not include a D band at 1350 cm$^{-1}$ (which would be indicative of defects), and the G:2D ratio is maintained in the pre- and post-pH jump Raman spectra[23,25]. Like the UV-Vis spectra, there was no appreciable change in the Raman spectra, confirming that the integrity of the graphene films is maintained at high and low pH conditions, and while subjecting the graphene samples to our experimental procedures, even for several days of SHG experiments.

**Supplementary Note 10. Assessment of Diffusion Kinetics in the Experimental Sample Cell.**
Similar to the analysis in our previous work[17], we wanted to test whether the changes in our SHG response were kinetically controlled rather than mass-transfer limited. First, we compared the duration of the SHG increase/decrease to the bulk diffusion time. To determine whether a concentration gradient would lead to significant bulk to surface diffusion times, we calculated the bulk to surface diffusion time according to

$$\tau_{diff} = \frac{\left(\frac{\Gamma_{1/2}}{C_{bulk}}\right)^2}{D_{bulk}} \qquad (3)$$

where $\tau_{diff}$ is the diffusion time from the bulk to the surface, $\Gamma_{1/2}$ is the absolute 50% saturation coverage, $C_{bulk}$ is the bulk proton concentration, and $D_{bulk}$ is the bulk proton diffusion coefficient in water[17,41]. The diffusion time, $\tau_{diff}$, as expressed above, is the time that it takes a proton to travel from the edge of the region, where 100% of the protons in solution are depleted in order to achieve 50% saturation surface coverage. As discussed above, approximately 25% of the silanol groups are deprotonated (SiO$^-$) at pH 10, and approximately 10% of the silanol groups are deprotonated at pH 3[9]. If we assume between $10^{15}$ and $10^{14}$ total surface sites cm$^{-2}$ [17,42], then at any point there are between ~ 1 x $10^{13}$ and 2.5 x $10^{14}$ surface sites that will undergo protonation or deprotonation. Given the bulk diffusion coefficients[43] of OH$^-$(aq), 5.30 x $10^{-5}$ cm$^2$ s$^{-1}$, and H$^+$(aq), 9.31 x $10^{-5}$ cm$^2$ s$^{-1}$, we obtain upper limits on the bulk to surface diffusion times ranging between 0.004 and 13.0 milliseconds, respectively. Under these timescales we can be confident that the processes sampled in our experiments are not diffusion limited.

In a second analysis, we calculated the thickness of the hydrodynamic boundary layer, which is defined the distance from a solid object (in this case, the wall of the flow tubes) to the location where the fluid velocity is 99% that of the bulk velocity[44]. Assuming laminar flow conditions the hydrodynamic boundary layer thickness can be calculated using the following expression

$$\delta = \left(\frac{v}{D}\right)^{-1/3} \sqrt{\frac{vx}{u_0}} \qquad (4)$$

where $\delta$ is the boundary layer thickness, $v$ is the kinematic viscosity of water at room temperature (0.009 cm$^2$ s$^{-1}$), D is the proton bulk diffusion coefficient, $x$ is the distance from the flow cell entrance to the focus spot of the laser (0.5 cm), and $u_0$ is the mean stream velocity[17,41]. Using a mean stream velocity of 1.1 cm s$^{-1}$ (~1 cm inner diameter tubing, 0.9 mL s$^{-1}$ flow rate), $\delta$ is approximately 130 $\mu$m. Since the hydrodynamic boundary layer thickness here is much smaller than a theoretical 1 cm diameter pellet, curvature effects can be neglected and the mass transfer coefficient, k$_c$, can be expressed as a function of the mass stream velocity alone using the Frössling correlation[44]:

$$k_c = \frac{D}{d_p}\left(2 + 0.6\sqrt{Re}^3 \sqrt{Sc}\right) \qquad (5)$$

Here D is the bulk proton diffusion coefficient, $d_p$ is the diameter of the pellet, $Re$ is the Reynolds number ($Re = u_0/vd_p$), and $Sc$ is the Schmidt number ($Sc = v/D$)[17,41,44]. In this simplified expression the mass transfer coefficient, $k_c$, depends on the mean stream velocity, $d_p$[17]. As shown in Supplementary Table2, we determined that the proton adsorption rate was independent of flow velocity. This assessment agrees with our previous report using the same experimental setup as discussed here, where we determined that the measured ion adsorption rate and overall SHG intensity was independent of the flow velocity[17]. Given that the mean stream velocity is independent of proton adsorption, we are confident that the acid/base reactions occurring at the fused silica surface are not mass transfer limited.

**Supplementary Note 11. Consideration of Vacancy Reconstruction.**

The formation of defect sites formed in the absence of water or oxygen during graphene synthesis lead to reconstruction of the carbon structure resulting in a more stabilized Stone-Wales (SW) type defects to eliminate the unsaturated sites as was found by Büttner et al.[45]. Our own DFT simulations carried out in vacuum show similar results in that the most stable defect sites in vacuum are those in which the unsaturated carbon sites rearrange to form 5 member aromatic ring structures to form double bonds to coordinatively saturate the carbon sites (Supplementary Fig. 17).

DFT calculations carried out by us further show that the unsaturated carbon defect sites can form covalent C-O or C-OH bonds in the presence of oxygen or water and as such spontaneously ring open the 5, 7 and 8 member ring structures to form the OH-terminated 4V site discussed in our present work. Specifically, DFT calculations indicate that while the SW type defect is the most stable configuration in the absence of water or oxygen, it ring opens as hydroxyl groups are brought in contact with the carbon atoms of the defect site. The defect site spontaneously ring opens to form the more favorable 4V sites in which 6 OH groups terminate the unsaturated carbon centers (Supplementary Fig. 18). This result suggests that the reconstructed SW-type defect, which is favorable under dry (vacuum) conditions, likely ring opens to form 4V defect in aqueous environments.

ReaxFF simulations were also carried out to examine the lowest energy states for the quad vacancy (4V) defects under dry conditions upon termination by surface hydroxyl intermediates as is shown in Supplementary Fig. 19a. ReaxFF-based energy-minimization calculations were carried out with a convergence criterion of 0.25 kcal Å$^{-1}$ to obtain the binding energy per carbon atom for the 4V and reconstructed 4V defect (R4V). The binding energy per carbon atom for the 4V-case is 178.45 kcal mol$^{-1}$ and for the R4V case it is 180.20 kcal mol$^{-1}$. As such, the R4V system is energetically more stable as compared to the 4V system, in agreement with the results from Buttner and co-workers[45].

We proceed to consider 6 hydroxyl (6OH) terminations of the 4V and R4V defects since the defects will be functionalized in presence of water. The binding energy per atom for the 4V+6OH case is 171.03 kcal mol$^{-1}$ and for the R4V+6OH case is 169.18 kcal mol$^{-1}$. We equilibrated the R4V+6OH system in the NPT ensemble for 25 ps with a time step of 0.25 fs using the Berendsen thermostat with a coupling time constant of 100 fs, and Berendsen barostat with a coupling time constant of 500 fs to control temperature and pressure of the entire system. We observe that the 5-membered rings at the R4V defect open up and the R4V+6OH system relaxes to the unreconstructed 4V+6OH system (Supplementary Fig. 19b). This result implies that upon functionalization of the defect sites, the reconstructed 4V defect relaxes to the unreconstructed 4V case considered in our MD simulations.

**Supplementary Note 12. Validation of ReaxFF for Water/Graphene Interfaces.**

To validate the ReaxFF graphene/water force field, we compare results from our ReaxFF molecular dynamics simulations of our water on graphene interface (Supplementary Fig. 20a) to long-time *ab initio* molecular dynamics (AIMD) simulations. Supplementary Fig. 20b and 20c show the number density profile of oxygen and hydrogen, comparing AIMD results with our ReaxFF force field. While the strength of the first peak is significantly higher using ReaxFF, the position of the first peak compares very well between the two methods. In addition, the $n_H(r)$ profile shows the presence of a ~1Å hydrophobic gap, in good agreement with AIMD as well as existing experiments.[46,47] The peak is around ~2.4Å from the graphene surface using both AIMD and ReaxFF, suggesting that the water orientation is captured properly using ReaxFF at the interface. Beyond ~7Å, water fluctuates about its bulk density. AIMD was performed using the VASP[48] code and a PBE exchange correlation functional with Grimme parameterization to describe the van der Waals interaction.[49] The AIMD was performed at T=300K, with a time step of 0.5 fs, for 43 ps. ReaxFF molecular dynamics were performed for over 150 ps at T=300K. All simulations are for an NVT ensemble at a density of 1g cm$^{-3}$ with a hundred water molecules.

**Supplementary Note 13: Proton diffusion through ether/pyrylium-terminated quad-defect sites and those terminated with OH groups.**

The O-terminated defect sites on graphene have been suggested to be in the form of ethers, carbonyls, and lactones [50,51]. The O atom can also sit at the carbon vacancy sites to form cationic aromatic pyrylium ($C_5H_5O^+$) species. DFT optimizations carried out for the O atom at the edge of the vacancy found that oxygen readily substitutes for C atoms to form an aromatic pyrylium species as is shown in Supplementary Fig. 13 and Supplementary Fig. 21. The O atom sits directly in the graphene plane where the C=O bond and C=C bond lengths are calculated to be 1.33 and 1.37 Å, respectively, characteristic of the aromatic C=O and C=O groups. The charges that were calculated indicate that the $C_5H_5O$ species is positively charged, which is fully consistent with pyrylium cation intermediates.

The three bridging O species that form at the 4V site sit directly within the graphene plane and form pyrylium intermediates (Supplementary Fig. 21). The hydrophobic character of the graphene surface, as well as the cationic charge on the pyrylium species, make it very difficult to transfer protons to the defect oxygen centers. While proton transfer readily occurs in solution phase above and below the graphene surface with barriers of only 0.2 eV, the activation energy required to diffusion through the 3O-terminated 4V defect site was calculated to 1.8 eV (Supplementary Fig. 22 and Table 1) and very likely does not occur at the low and moderate temperatures in this system. Further simulations were carried out to examine proton diffusion through the same 3O pyrylium functionalized defects at the water/graphene/water/$SiO_2$ interface to examine the role of the underlying $SiO_2$. The calculated barrier (Supplementary Fig. 23) was found 2.5 eV, which is slightly higher than that for the water/graphene water interface. The $SiO_2$ structure partially limits the mobility and freedom of the interfacial water, which acts to increase the activation energy.

Hydroxyl terminated vacancy sites can readily form in the presence of water and result in effective hydrogen bonding networks that can stabilize protons and provide efficient conduits for proton transfer. The 6 unsaturated carbon sites in the 4V site react with water to form 6 terminal

C-OH groups. The most stable conformation of the 6 OH groups is one in which three of the hydroxyl groups are oriented into the solution phase above the surface and three are oriented into the solution below the surface (see Supplementary Fig. 24). The strong hydrogen bonds that form between the surface hydroxyl groups and water molecules stabilize protons and provide flexible pathways for proton transfer into solution.

Our DFT calculations show that protons in solution diffuse through the hydroxylated vacancy site with a barrier of only 0.68 eV (see Supplementary Fig. 25 and Table 1). This process can readily proceed at room temperature. The protons are transferred via proton shuttling following a Grotthuss mechanism[52] where the proton in solution take on the form of an $H_3O^+$ hydronium species which can very efficiently transfer its proton along a network of O-H-O-H bonds that make up the proton relay conduit from the water molecules above the surface through defect OH sites bound to the graphene surface and onto the water molecules below the surface thus resulting in a low activation barrier of only 0.68 eV. This was the most effective proton transfer path found.